%% file: zingaro-tavr-arxiv.tex
\documentclass[12pt]{article}
\usepackage{authblk}
\usepackage{booktabs}
\usepackage{url}
\usepackage{hyperref}

\usepackage[margin=2cm]{geometry}
\usepackage{comment}
\usepackage[numbers]{natbib}

\include{mycommands}

\begin{document}

\title{Advancing aortic stenosis assessment: validation of fluid-structure interaction models against 4D flow MRI data}
\author[1,*]{Alberto Zingaro}
\author[1,*]{Irmantas Burba}
\author[1]{David Oks}
\author[1]{Mauro Fontana}
\author[2]{Cristóbal Samaniego}
\author[3]{Micha Bischofberger}
\author[4]{Bart de Boeck}
\author[3]{Andre Douverny}
\author[3]{Özge Karakas}
\author[4]{Stefan Toggweiler}
\author[5, 6]{Dabit Arzamendi-Aizpurua}
\author[3]{Utku Gülan}
\author[1, 2]{Mariano Vázquez}

\affil[1]{\small{ELEM BioTech,
            {Pier 07, Calle Laietana, 26}, 
            {Barcelona},
            {08003}, 
            {Spain}}}

\affil[2]{\small{Barcelona Supercomputing Center,
            {Plaça d'Eusebi G\"uell, 1-3}, 
            {Barcelona},
            {08034}, 
            {Spain}}}

\affil[3]{\small{Hi-D Imaging,
            {Technoparkstrasse 2}, 
            {Winterthur},
            {8406}, 
            {Switzerland}}}

\affil[4]{\small{Lucerne Cantonal Hospital (Luzerner Kantonsspital),
            {Spitalstrasse}, 
            {Luzern},
            {6000}, 
            {Switzerland}}}

\affil[5]{\small{Cardiology Department, Hospital de la Santa Creu i Sant Pau.,
            {Carrer St. Antoni M. Claret 89}, 
            {Barcelona},
            {08025}, 
            {Spain}}}

\affil[6]{\small{Department of Medicine, Universitat Autònoma de Barcelona,
            {Pg. Vall d'Hebron, 119-129. }, 
            {Barcelona},
            {08023}, 
            {Spain}}}
\affil[*]{\footnotesize{These authors equally contributed to this work}}

\date{}

\maketitle

\begin{abstract}
Systematic \textit{in vivo} validations of computational models of the aortic valve remain scarce, despite successful validation against \invitro{} data. Utilizing a combination of computed tomography and 4D flow magnetic resonance imaging data, we developed patient-specific fluid-structure interaction models of the aortic valve immersed in the aorta for five patients in the pre-transcatheter aortic valve replacement configuration. Incorporating also an \invitro{} setup of the valve, our computational models are subjected to rigorous validation against 4D flow measurements. Our results demonstrate the models' capacity to accurately replicate flow dynamics within established ranges of uncertainties mainly arising from 4D flow noise. In addition, we illustrate how computational models can serve as valuable cross-checks to reduce noise and erratic behaviour of \invivo{} data. This study represents a significant step towards integrating \insilico{} technologies into real clinical contexts, providing a robust framework for improving aortic stenosis diagnosis and the design of next-generation aortic valve bioprostheses. 
\end{abstract}


\section{Introduction}

\acrfull{as} represents a constriction of the \acrfull{av} opening and stands as the most prevalent valvular pathology in developed nations, affecting 9 million people globally. Left untreated, it dramatically increases the risk of sudden cardiac death and heart failure \cite{aluru2022valvular, aha_as} with a one-year mortality rate as high as 50\%.  \acrfull{tavr} addresses \acrshort{as} by replacing a dysfunctional or diseased \acrshort{av} with a prosthetic valve made from animal pericardial tissue \cite{nih_tavr}. Distinguished from conventional methods like \acrlong{savr}, \acrshort{tavr} emerges as a minimally invasive alternative, offering distinct advantages for patients \cite{cribier2002percutaneous, fanning2013transcatheter}. The severity of \acrshort{as} and the outcomes of \acrshort{tavr} are typically gauged using various \acrfull{qois} such as \acrfull{ava}, \acrfull{tpg}, peak transvalvular velocity, velocity ratio, and \acrfull{lvot} diameter. These \acrshort{qois} are routinely assessed using imaging techniques such as doppler echocardiography, \acrfull{cmr}, \acrfull{ct} angiography, or more advanced techniques such as 4D flow \acrfull{mri}. \cite{baumgartner2009echocardiographic, otto1986determination, zoghbi2009recommendations, archer2020validation}. 
However, these techniques have limitations. The reliability of the assessment is intricately tied to the operator's experience and expertise, thereby introducing potential variability \cite{manzo2023echocardiographic}. Furthermore, challenges arise in effectively visualizing specific flow structures -- particularly in scenarios involving eccentric jets and turbulent flows -- thereby contributing to suboptimal image quality and errors in velocity measurements \cite{manzo2023echocardiographic, ngo2019four}.  
Additionally, in cases of low ejection fraction, further limitations may be encountered, potentially complicating the interpretation of results and accuracy of assessments.
{Advanced techniques like \fourd{}, despite their ability to visualize the velocity field, are predominantly limited to research applications rather than being integrated into clinical routine. } Additionally, the computation of certain \acrshort{qois} (such as the \acrshort{ava}) relies on formulas based on several hypotheses about blood flow  \cite{clavel2015aortic}. Consequently, both random and systematic measurement errors may potentially impact  \acrshort{as} assessment, possibly leading to misdiagnosis and subsequent mistreatment with \acrshort{tavr} \cite{velders2022measurement}.
The precision of estimating these \acrshort{qois} can be heightened by complementing these imaging techniques with  \insilico{} technologies, including \acrfull{cfd} and \acrfull{fsi} \acrshort{av} models. The latter specifically entails the interaction between blood hemodynamics and the mechanical behavior of the \acrshort{av}, providing the computations of several \acrshort{qois} useful in the context of \acrshort{as} risk prediction and \acrshort{tavr} assessment \cite{bahraseman2016fluid, zhu2022computational, fumagalli2023fluid, luraghi2019modeling, luraghi2020impact, ghosh2020numerical, basri2020fluid, de2003computational, oks2023effectsino, oks2023effect, oks2022fluid}.

A wealth of literature exists on mathematical models of 
cardiac valves and their interaction with cardiac hemodynamics \cite{mcqueen1982fluid, de2003computational, astorino2009fluid, griffith2012immersed, marom2015numerical,  aboelkassem2015mathematical, flamini2016immersed, fedele2017patient,  spuhler20183d, wu2019immersogeometric, fumagalli2020image, oks2022fluid, zingaro2022modeling, bennati2023turbulent, zingaro2024electromechanics} and for a comprehensive review of \acrshort{fsi} models for the blood-\acrshort{av}, we refer to the recent paper by Kuchumov et al. \cite{kuchumov2023fluid}. 
However, for \textit{credibility evidence} in computational modeling for medical devices submission, % as pointed out in the last report by the \acrfull{fda},
different \acrfull{vvuq} activities should be carried out, including validation processes against \invitro{} and \invivo{} datasets \cite{fda_vvuq}. Several studies have presented computational \acrshort{fsi} models of the blood-\acrshort{av} system conducting validation against \invitro{} data. {\textit{In vitro} flow measurements have been utilized to extract hemodynamics parameters in anatomically accurate silicone models and to validate numerical models in the literature \cite{gallo2014analysis, toggweiler2022turbulent, gulan2022comparative}. } 
 Dumont et al. carried out a validation study of the blood-\acrshort{av} \acrshort{fsi} in a 2D configuration against \invitro{} data \cite{dumont2004validation}. Luraghi et al. compared \acrshort{fsi} and structural mechanics simulations of polymeric heart valves against \invitro{} data, showing the capability of \acrshort{fsi} in the replication of different patterns of the experiments \cite{luraghi2017evaluation}. In 2018, Tango et al presented a 3D validation study aimed at replicating an \invitro{} \acrfull{piv} setup of the \acrshort{av} \cite{tango2018validation}.  Sodhani et al. carried out a validation study against \invitro{} data for an artificial textile reinforced \acrshort{av} \cite{sodhani2018fluid}. 
In the context of \acrshort{tavr}, Borowski et al. \cite{borowski2022validation} validated an \acrshort{fsi} model against \acrshort{piv} data. Thus, the literature on \acrshort{fsi} models of \acrshort{av} dynamics is extensive, showcasing the robust theoretical foundation of these computational models. 

Furthermore, during the past years -- in line with the \textit{credibility evidence} activities pointed out by the \acrfull{fda} and other regulatory agencies -- particular attention has been devoted to the validation of these computational models, always in the context of replicating experimental setup. However, it is noteworthy that none of these models have undergone systematic validation against \invivo{} data, which somewhat still constrains their application in real clinical contexts. Indeed, the validation of computational models in cardiovascular applications against \invivo{} data poses a formidable challenge. This challenge arises primarily from the limited availability of data crucial for configuring boundary conditions and selecting specific model parameters (such as in the constitutive law for valve mechanics). Additionally, medical data often come with substantial noise and error, and their precision frequently hinges on the expertise of the individuals acquiring these measurements. This further complicates the process of data assimilation in the computational model and renders it challenging to assess the accuracy of numerical results in comparison to these inherently noisy \invivo{} datasets. This gap in the literature underscores an urgent need for the validation of \acrshort{fsi} models of the \acrshort{av} against \invivo{} datasets. 

The aim of this paper is to present a pipeline for the validation of \acrshort{fsi} computational model of the  \acrshort{av} against \invivo{} measurements. Utilizing a combination of \acrshort{ct} and \fourd{} data, we build a computational model representing the \acrshort{av} immersed in the aorta for five patients in the pre-\acrshort{tavr} configuration (i.e. prior to \acrshort{tavr} procedure). Our computational model employs the \acrfull{ib} method \cite{peskin1972flow, griffith2012immersed} for \acrshort{fsi}, as we recently proposed in \cite{oks2022fluid}, and undergoes validation against \fourd{} measurements. 
We establish the credibility of our model through a comprehensive comparison of various indicators and flow patterns. We demonstrate the computational model's ability to accurately replicate flow dynamics, both qualitatively and quantitatively. Additionally, we demonstrate how our \acrshort{fsi} computational model serves as a cross-check to mitigate noise and erratic behaviour in \invivo{} data. We discuss the pivotal role of numerical models and simulations in the context of \acrshort{as}, serving as tools capable of overcoming some of the limitations associated with current \invivo{} assessment, and enabling a more comprehensive understanding of complex physiological phenomena as \acrshort{av} dynamics. To the best of our knowledge, this is the first work that validates \acrshort{fsi} computational models of the \acrshort{av} against \invivo{} data for several patients, marking a significant advancement in the integration of \insilico{} technologies in real clinical contexts. 

This paper is organized as follows: \Cref{sec:methods} introduces patient-specific data, the pipeline we developed to process them, our computational model and the validation data from \fourd{} acquisitions. In \Cref{sec:results}, we present our computational results against the \invivo{} dataset and additional \acrshort{qois} that can be computed with \acrshort{fsi} simulations. \Cref{sec:discussion} is devoted to the discussion and we present limitations and conclusions in \Cref{sec:limitations} and \Cref{sec:conclusions}, respectively.

\section{Methods}
\label{sec:methods}

We present the methodology of the work. In \Cref{sec:methods-patients-selection} we introduce the patients cohort. \Cref{sec:methods-ct-mri-postpro} is focused on the postprocessing of \acrshort{ct} and \fourd{} data. In \Cref{sec:methods-invitro}, we introduced the \invitro{} setup from which we get the \acrshort{av} geometry. In \Cref{sec:methods-fsi}, we show the computational model and its setup, whereas we explain how we feed the model with the \invivo{} flowrate in \Cref{sec:methods-data-assimilation-flowrate}. In \Cref{sec:results-calibration-young} we present a calibration study aimed at selecting the valve's Young modulus.   

\subsection{Patients selection}
\label{sec:methods-patients-selection}

% P1 : TT07
% P2 : TT04
% P3 : TT10
% P4 : TT03 
% P5 : TT05 

\begin{table}
	\centering
	\begin{tabular}{cccccc}
		\toprule
		& P1 & P2 & P3 & P4 & P5\\
		\midrule 
		age & 81 & 77 & 89 & 74 & 66 \\
		sex & M & M & F & F & M \\
            height [\si{\centi\meter}]	& 182 & 169	& 155 & 157	& 175 \\
            weight [\si{\kilo\gram}]	& 78 & 64& 56& 46 &	130	\\
            annulus area [\si{\milli\meter\squared}] &	543&	530&	344&	454&	405 \\
            annulus perimeter [\si{\milli\meter}]	& 83	& 83	& 66	& 77	& 72 \\
		%\acrshort{as} severity? & ? & ? & ? & ? & ? \\
            $\Thb$ [$\si{\second}$] & \num{1.02} & \num{1.06} & \num{0.72} & \num{0.71} & \num{0.78} \\ 
		\bottomrule 
	\end{tabular}
	\caption{Patients information. $\Thb$ is the heartbeat period. }
	\label{tab:patients-info}
\end{table}
{We consider clinical data belonging to the TURBULENT TAVI study\footnote{``Correlation between bench and clinical turbulent and mean kinetic energy after transcatheter aortic valve Implantation of the Allegra valve: the TURBULENT TAVI pilot study'', Luzerner Kantonsspital, Luzern, PI: PD Dr. med. Stefan Toggweiler, project ID 2019-01344}, aiming to provide a better understanding of the intensity of turbulence in the ascending aorta before and after \acrshort{tavr}. The study hypothesizes that there is a correlation between \invitro{} and \invivo{} measurements. In the scope of the study, pre-\acrshort{tavr} and post-\acrshort{tavr} scans for patients undergoing \acrshort{tavr} at Heart Center Luzern, Switzerland were performed. }
In this paper, we consider five patients characterized by the same level of \acrshort{as} severity, namely P1, P2, P3, P4, and P5. Patient information is provided in \Cref{tab:patients-info}. For these patients, we have available \acrshort{ct} and \fourd{} acquisitions.

\subsection{\acrshort{ct} and \fourd{} post-processing}
\label{sec:methods-ct-mri-postpro}

\acrshort{ct} data are used to generate the volumetric fluid mesh, whereas \fourd{} is used for both data assimilation (in terms of input blood flowrate) and as a validation set (in terms of the whole velocity field and additional \acrshort{qois}).

\begin{figure}[t]
	\centering
	\includegraphics[trim = {0cm 0cm 0cm 0cm}, clip, width=\textwidth]{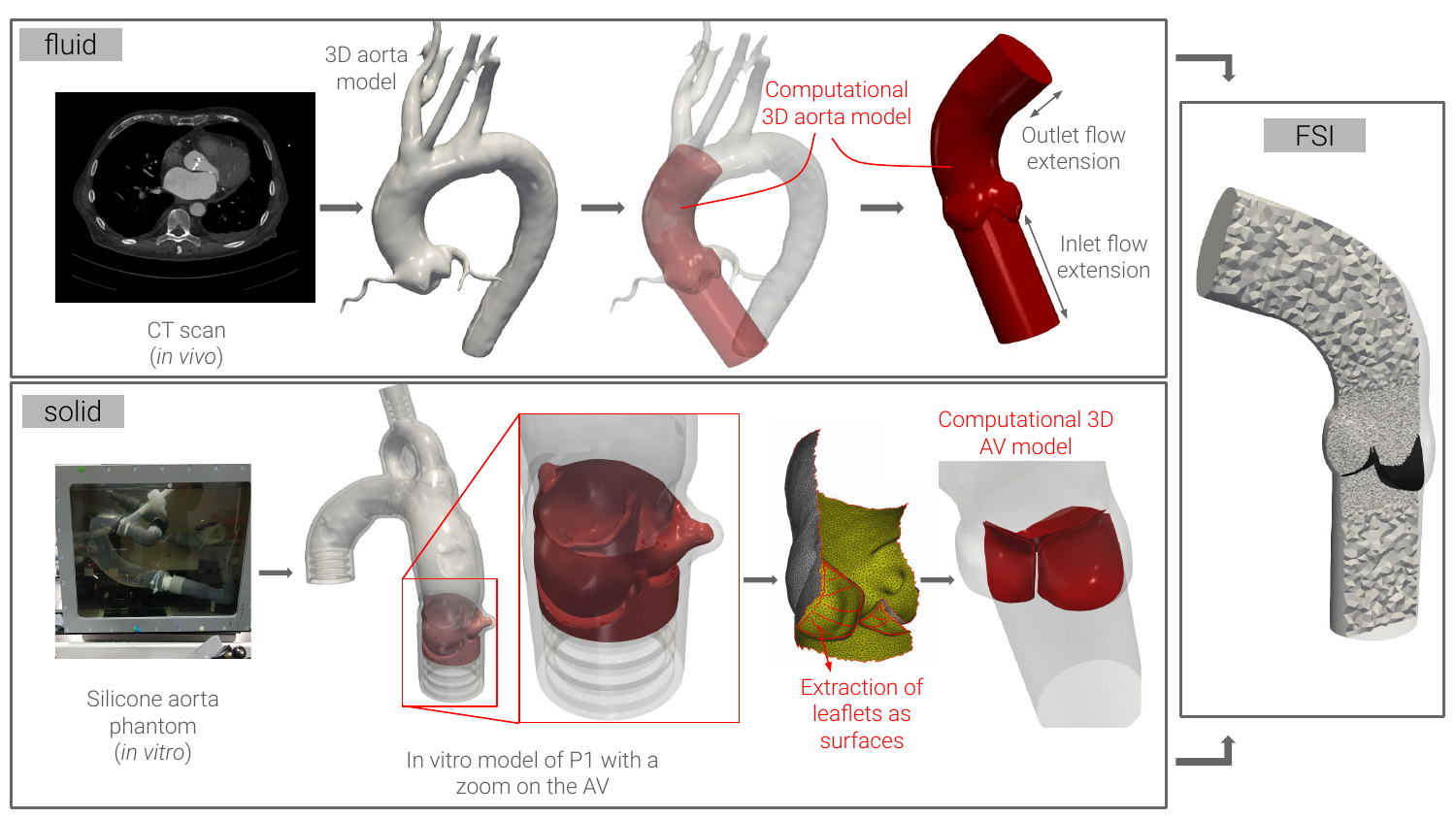}
	\caption{Preprocessing pipeline to generate the \acrshort{fsi} meshes. Top: pipeline for the aorta: \invivo{} CT scan acquisition of the aorta, 3D segmented aorta, coronary arteries clipping and flow extensions adding to get the final computational domain. Bottom:  preprocessing pipeline to generate our template \acrshort{av} mesh from the \invitro{} experiment of P1.}
		\label{fig:preprocessing-pipeline}
\end{figure}

\begin{figure}[t]
	\centering
	\includegraphics[trim = {0cm 0cm 0cm 0cm}, clip, width=\textwidth]{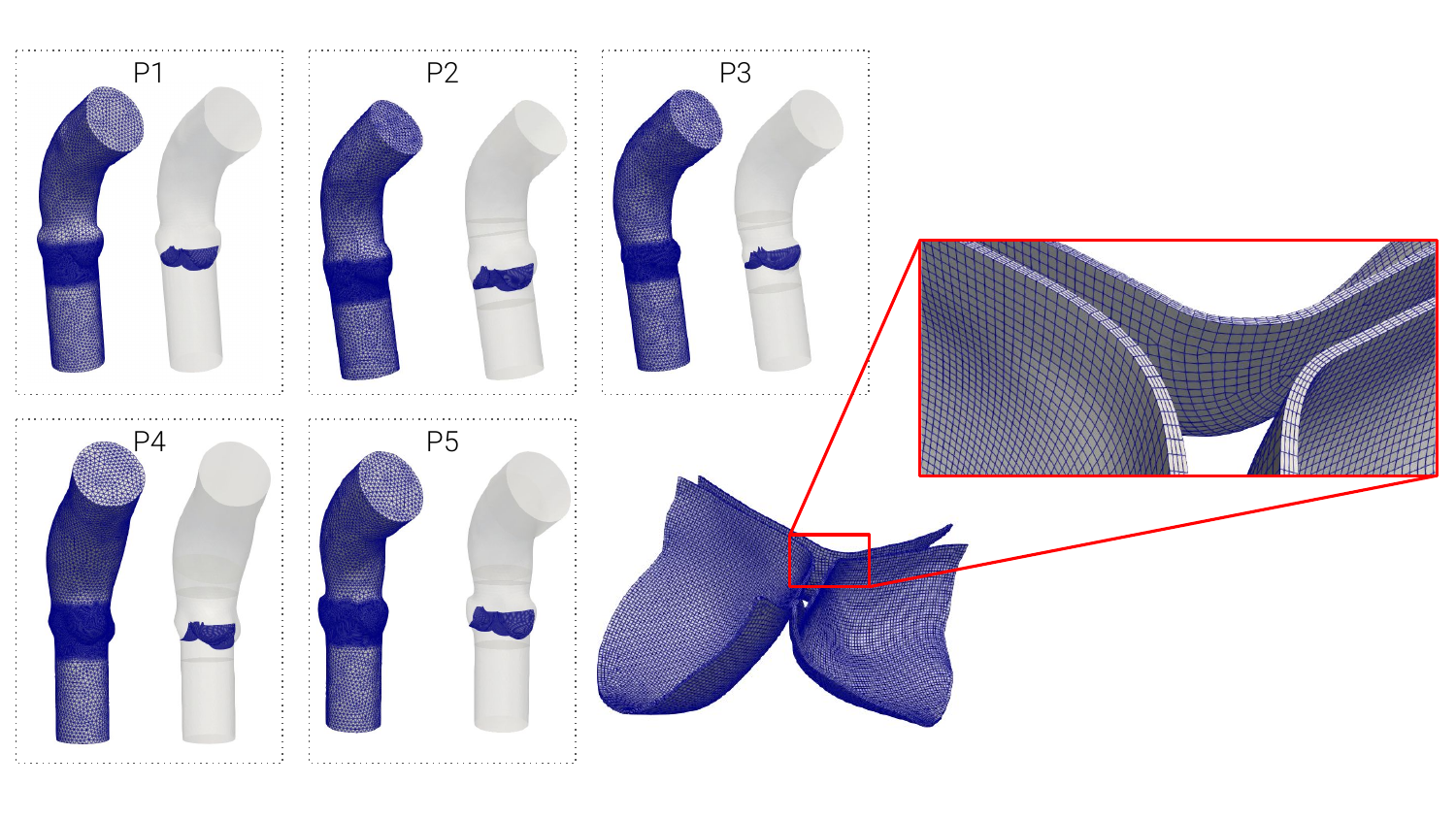}
	\caption{Computational meshes. Left: the five meshes of the aorta and \acrshort{av}; right: the \acrshort{av} solid mesh with a zoom on its thickness. }
	\label{fig:fluid-solid-meshes}
\end{figure}

\paragraph{The \acrshort{ct} data}
We segmented the aorta of the five patients using 3D Slicer \cite{kikinis20133d}, with particular emphasis on cleaning the leaflets region. Then, we perform some preprocessing steps to get the final mesh, as we show in \Cref{fig:preprocessing-pipeline}. These include a clipping of the original aortic geometry to include only the ascending portion of the aorta. Moreover, we remove the coronary arteries and we add flow extensions to both inlet and outlet sections. This is done to allow the fluid to fully develop as it enters, and to avoid potential instabilities at the outlet section.

The preprocessing of all the geometries and the creation of the meshes is carried out with the preprocessing software ANSA \cite{ansa}. We generate tetrahedral meshes of the aorta, as we showcase in \Cref{fig:fluid-solid-meshes} (left). Notice that, for the \acrshort{fsi} model to properly capture the \acrshort{av}, we conveniently refine the fluid mesh in a region closer to the valve. Details of the generated meshes are provided in \Cref{tab:meshes} (top).

\begin{table}
	\centering
	\begin{tabular}{llccccc}
		\toprule
            & & \multicolumn{5}{c}{\textbf{fluid}} \\
            \textbf{patient} & & P1 & P2 & P3 & P4 & P5 \\
            %\midrule
            \cmidrule(lr){3-7}
            \textbf{mesh type} & & \multicolumn{5}{c}{tetrahedra} \\
            \h{min} & [\si{\milli\meter}] & \num{0.24} & \num{0.14} & \num{0.15}& \num{0.24} & \num{0.25}  \\
            \h{avg} & [\si{\milli\meter}] & \num{0.61}  & \num{0.45} & \num{0.46} & \num{0.56} & \num{0.57}\\
            \h{max} & [\si{\milli\meter}] & \num{2.58} & \num{2.11}& \num{2.53} & \num{2.55} & \num{2.49} \\
            \textbf{\# of elements} & & \num{1381518}  & \num{1409841} & \num{977630}& \num{1309822}& \num{1141088} 
            \\
            %total \acrshort{dofs} & & \num{2324700} & \num{1373680}& \num{995980} & & & \\
		\bottomrule
            & &  & & & & \\
		\toprule
            & & \multicolumn{5}{c}{\textbf{solid}} \\
            \textbf{patient} & & P1 & P2 & P3 & P4 & P5 \\
            %\midrule
            \cmidrule(lr){3-7}
            \textbf{mesh type} & & \multicolumn{5}{c}{hybrid hexahedral/pentahedral} \\
            \h{min} & [\si{\milli\meter}]  & \num{0.0480}& \num{0.0742} & \num{0.0750}  & \num{0.0735} & \num{0.0730} \\
            \h{avg} & [\si{\milli\meter}] &   \num{0.0759} & \num{0.0752}& \num{0.07502} & \num{0.0745} & \num{0.0747}\\
            \h{max} & [\si{\milli\meter}] & \num{0.1127}& \num{0.0747}& \num{0.0755} & \num{0.07515}& \num{0.0760} \\
            \textbf{\# of elements} & & \num{61032} & \num{58944} & \num{50848} & \num{46892} & \num{57120}
            \\
            %total \acrshort{dofs} & & \num{2324700} & \num{1373680}& \num{995980} & & & \\
		\bottomrule 
	\end{tabular}
	\caption{Details of the generated meshes: \h{min}, \h{avg}, and \h{max} are the minimum, average, and maximum mesh element sizes, respectively. Top: mesh for the aorta; bottom: mesh for the valve leaflets. } 
		\label{tab:meshes}
\end{table}

\paragraph{The \fourd{} data}

\begin{figure}[t]
	\centering
	\includegraphics[trim = {0cm 2.5cm 0cm 0cm}, clip, width=\textwidth]{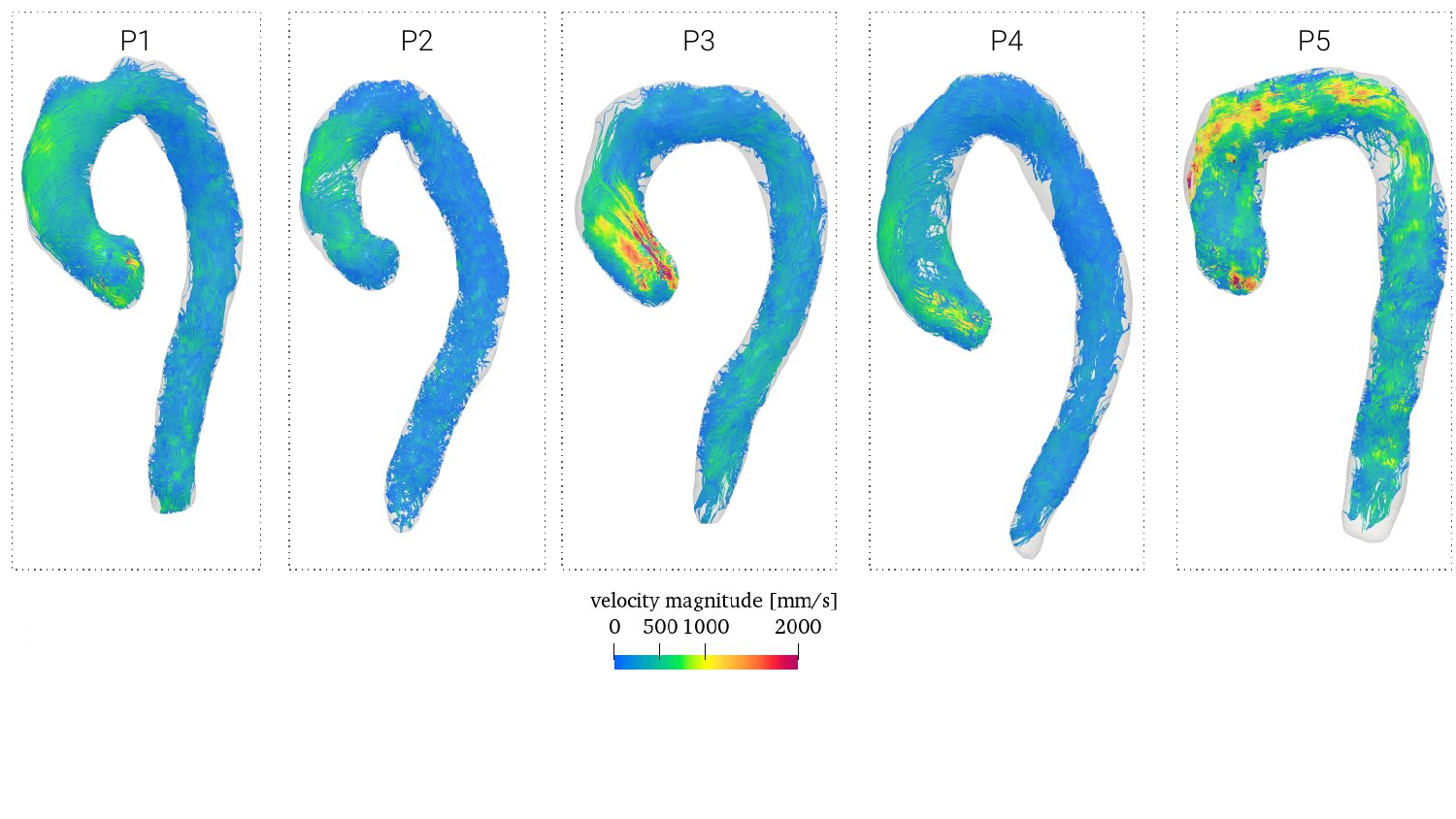}
	\caption{Streamlines obtained with \fourd{} at systolic peak for the five patients considered in the study.}
	\label{fig:4dflow-streamlines}
\end{figure}

{In vivo free-breathing 4D-MRI measurements were performed with a \SI{1.5}{\tesla} scanner (Siemens Magnetom Aera, Siemens AG, Munich, Germany) to extract the velocity field phase-averaged over subsequent cardiac cycles: $\langle \bm u ^\mathrm{4Dflow} (\bm x, t)\rangle$. The imaging parameters were the following: spatial resolution $2.5 \times 2.5 \times 2.5$ \si{\milli\meter\cubed}, field of view $250 \times 160 \times 50$ \si{\milli\meter\cubed}, TR/TE 5, 8/3, 4 \si{\milli\second}, and temporal resolution \SI{35}{\milli\second}. The velocity encoding values were 40, 100, and 200 \si{\centi\meter\per\second} per direction. The phase contrast sequences were synchronized with the electrocardiography. A 3D-gradient echo phase contrast sequence with multi-point velocity encoding was applied to obtain 10x $k-t$ undersampled data.} We give a graphical representation of the \fourd{} data in \Cref{fig:4dflow-streamlines}. We use the \fourd{} for data assimilation and as a validation set. Specifically, the \invivo{} flowrate serves as our input flowrate, whereas several quantities are used to validate our computational results. We better detail both processes in \Cref{sec:methods-data-assimilation-flowrate} and \Cref{sec:results}. 

\subsection{In vitro setup}
\label{sec:methods-invitro}

To construct the computational \acrshort{av} model, we leverage an \invitro{} setup meticulously crafted by \textit{Hi-D Imaging} using 3D printing technology\cite{gulan2017shear}. The experimental setup replicates the hemodynamics observed in P1 and is initiated from the same \invivo{} segmentation \cite{zeugin2024vitro}. 

We used the \invitro{} \acrshort{av} as a template valve for all the patients (by appropriately adapting it to the patient-specific aorta geometry). We have adopted a consistent \acrshort{av} template across multiple patients driven by the substantial time investment required for individualized valve mesh generation. Importantly, our decision is grounded in the consideration that the reference \fourd{} data available for our study are consistently acquired in the region downstream of the valve. Consequently, our ability to thoroughly validate the numerical solution in the immediate proximity of the valve is inherently limited.

To produce a generic valve geometry suitable for the whole patient cohort the following procedure -- illustrated in \Cref{fig:preprocessing-pipeline} -- is used. We start by extracting the inferior part of the aortic leaflets (i.e. the parts towards the left ventricle) from the 3D model of the \invitro{} setup, obtaining three surfaces. Then, we extend their fixed edges until they perfectly adhere to the computational fluid domain boundaries of P1 (obtained from \invivo{} data). Once the three surfaces are adapted to the aortic wall of the fluid domain, we extrude the leaflets in the normal direction to create a volumetric mesh with a constant thickness equal to \SI{0.3}{\milli\meter} \cite{thickness_av}. %This valve serves as the template geometry that we use for all the patients. 
Since this valve corresponds to patient P1, we conveniently adapt it to be used also with the remaining patients: we position the valve in the region of the \acrfull{sov}, we do a rotational alignment to the coronaries, and we scale the geometry according to the diameter of the patient-specific aorta considered. 

As for the aorta, the preprocessing of the valve is carried out with the software ANSA \cite{ansa}: we produce a hybrid hexahedral/pentahedral mesh using four elements in the thickness to capture the bending modes of the valve \cite{oks2022fluid}. The \acrshort{av} mesh is shown in \Cref{fig:fluid-solid-meshes} and details of the generated meshes for the solid domains are provided in \Cref{tab:meshes}.

\subsection{Computational model setup}
\label{sec:methods-fsi}

\begin{figure}[t]
\stackinset{l}{60pt}{t}{20pt}{\Large $\boxed{\omegaf}$}{%
\stackinset{r}{110pt}{t}{20pt}{\Large $\boxed{\omegas}$}{%
\stackinset{l}{60pt}{b}{15pt}{\Large $\gammafin$}{%
\stackinset{l}{55pt}{t}{70pt}{\Large $\gammafw$}{%
\stackinset{l}{200pt}{t}{20pt}{\Large $\gammafout$}{%
\stackinset{r}{164pt}{t}{165pt}{\Large $\swarrow$}{%
\stackinset{r}{120pt}{t}{120pt}{\Large $\gammaswet$}{%
\stackinset{r}{180pt}{t}{180pt}{\Large $\gammasw$}{%
\includegraphics[width=\textwidth]{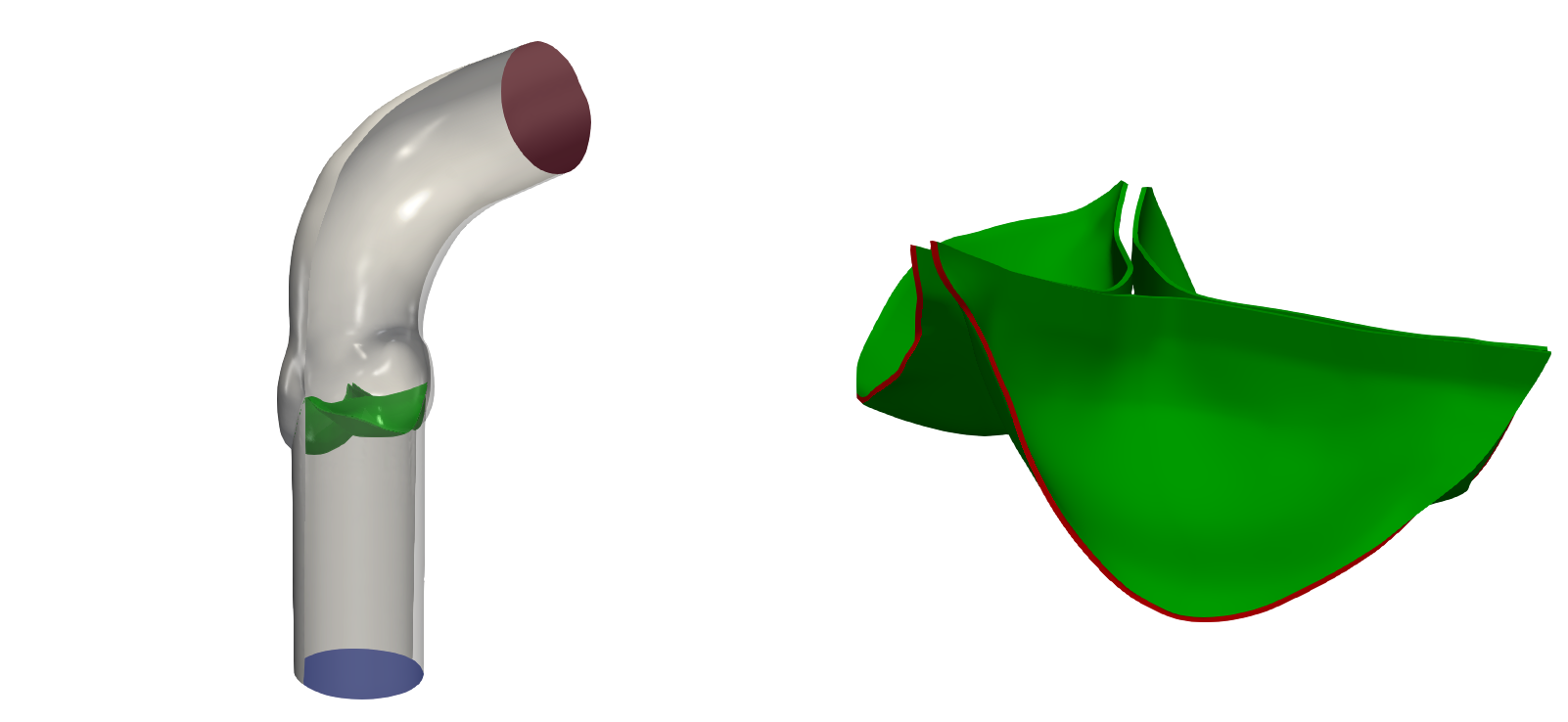}
}}}}}}}}
\caption{Computational domain. Left: the fluid domain $\omegaf$ bounded by inlet section $\gammafin$, outlet section $\gammafout$ and wall $\gammafw$. Right: the solid domain $\omegas$ bounded by $\gammasw$ and $\gammaswet$. }
\label{fig:domains}
\end{figure}

% Domains
Let $\omegaf$ and $\omegas$ be the fluid and solid domains, representing an ascending aorta and an \acrshort{av}, respectively (see \Cref{fig:domains}) and let $T$ be the final time. The fluid domain is bounded by $\gammafin$, $\gammafout$, and $\gammafw$, being the inlet section, outlet section, and wall, respectively. The solid domain is bounded by $\gammaswet$ and $\gammasw$, being the ``wet'' part of the aortic leaflets and the portion of the \acrshort{av} attached to the endothelium of the aorta, respectively.  

\paragraph{The fluid model}
We consider the blood as an incompressible, viscous, and Newtonian, fluid with constant density $\rhof = \SI{1100}{\kilo\gram\per\meter\cubed}$ and constant dynamic viscosity $\muf=\SI{3.6e-3}{\pascal\second}$. Let $\uf$ and $\pf$ be the fluid velocity and pressure, $\stressf(\uf, \pf) = - \pf \mathrm{I} + \muf \left ( \grad \uf + \grad^T \uf \right)$ the total stress tensor, the incompressible Navier-Stokes equations endowed with boundary and initial conditions read:
\begin{align}
    & \rhof\left(\pdv{\uf}{t} + \left(\uf \cdot\grad\right) \uf\right) - \div \stressf(\uf, \pf) = \forcing & \text{in }  \omegaf \times (0, T) , \label{eq:ns-momentum} \\
    & \div \uf = 0 & \text{in }  \omegaf \times (0, T) , \label{eq:ns-continuity} \\
    & \uf = \ufin & \text{on }  \gammafin \times (0, T) , \label{eq:ns-dirichlet-in} \\
    & \uf = \bm 0 & \text{on }  \gammafw \times (0, T) , \label{eq:ns-dirichlet-w}  \\ 
    & \stressf(\uf, \pf) \nf = \bm 0 & \text{on }  \gammafout \times (0, T) , \label{eq:ns-neumann-out} \\
    & \uf = \bm 0 & \text{in }  \omegaf \times \{ 0 \} . \label{eq:ns-ic}  
\end{align}
\Cref{eq:ns-momentum} is the momentum balance, with $\forcing$ a forcing term used to enforce dynamic balance according to the \acrshort{ib} method (see below). \Cref{eq:ns-continuity} is the continuity equation. At the inlet section, we set the Dirichlet boundary condition (\Cref{eq:ns-dirichlet-in}) by prescribing the inlet flowrate in time and a parabolic profile in space:
\begin{equation*}
    \ufin (\x, t) = - 2 \frac{\Qin(t)}{|\gammafin|} 
            \left( 1 - \frac{r^2(\x)}{{\Rin} ^2}
            \right ) \nf(\x).
\end{equation*}
where $|\gammafin|$ is the measure of $\gammafin$, $r = |\x|$ is the polar radius coordinate, $\x$ the space coordinate, $\Rin$ the radius of the inlet section, and $\nf$ its outward pointing normal. We denote by $\Qin$ the inlet flowrate computed with \fourd{}, as we explain in \Cref{sec:methods-data-assimilation-flowrate}. Notice that a parabolic profile would in principle correspond to a fully laminar flow. However, we create a flow extension in the inlet long enough so that the flow has space to develop, and our choice of the inlet spatial profile becomes negligible from the region immediately upstream of the \acrshort{av}. On the wall, we set no-slip condition (\Cref{eq:ns-dirichlet-w}) and we prescribe homogeneous Neumann boundary condition on the outlet section (\Cref{eq:ns-neumann-out}). We start the fluid problem with a null initial condition (\Cref{eq:ns-ic}). 

To account for possible transition-to-turbulence effects in the presence of \acrshort{as} \cite{stein1976turbulent}, we consider a \acrfull{les} model using the eddy viscosity subgrid-scale model proposed by Vreman et al. \cite{vreman2004eddy} and used in the vascular context in \cite{pal2014large, katz2023impact, oks2023effectsino}. Particularly in the context of \acrshort{as}, using a turbulence model might be imperative to capture the small scales when the orifice is small and high-speed jets  with large Reynolds numbers are computed \cite{manchester2021analysis, oks2023effectsino}. 

% Solid
\paragraph{The solid model}
We model the \acrshort{av} dynamics with the elastodynamic equations in a total Lagrangian formulation. Let $\omegaszero$ be the solid domain in its reference (initial) configuration and let $\X$ be the position vector w.r.t. $\omegaszero$. We define $\chi(\X, t)$ as the mapping from the reference to the current configuration so that the displacement of the solid is $\displ(\X, t) = \chi(\X, t) - \X$. Let $\rhos$ and $\rhoszero$ be the solid density in the current and in the reference configuration, $F = \pdv{\chi}{\X}$ the deformation tensor, and its determinant $J = \det(F)$. The elastodynamic equation endowed with boundary and initial conditions reads:
\begin{align}
& \rhoszero \pdv{^2 \displ}{t^2} - \divzero P(\displ) = \bm 0 & \text{ in } \omegaszero \times (0, T), \label{eq:solid-mom}\\
& \rhoszero J^0 = \rhos J & \text{ in } \omegaszero \times (0, T), \label{eq:solid-mass}\\
& \displ = \bm 0 & \text{ on } \gammasw \times (0, T), \label{eq:solid-dirichlet} \\
& \pdv{\displ}{t} = \bm 0 & \text{ in } \omegaszero \times \{0\}, \label{eq:solid-ic-ddot} \\ 
& P(\displ) = 0 & \text{ in } \omegaszero \times \{0\}, \label{eq:solid-ic-P}
\end{align}
where \Cref{eq:solid-mom} is the momentum balance, with $P$ the nominal stress tensor. \Cref{eq:solid-mass} is the mass conservation, we set a homogeneous Dirichlet condition on $\gammasw$ (\Cref{eq:solid-dirichlet}), and we initialize our problem with null initial conditions (\Cref{eq:solid-ic-ddot}, \Cref{eq:solid-ic-P}). The material of the \acrshort{av} is primarily characterized by an isotropic extracellular matrix and anisotropic alignment of collagen fibers. However, for simplicity, we adopt an isotropic Neo-Hookean material model with Lamé coefficients $\lambda_\mathrm{s}$ and $\mu_\mathrm{s}$ \cite{belytschko2014nonlinear, oks2022fluid}. Our choice of this simplified constitutive law is tailored to the specific objectives of our work, deemed adequate for validating blood flow downstream of the valve. Furthermore, our selection is in line with other studies in the literature \cite{spuhler20183d, de2003three, oks2022fluid}.

% Interface conditions and IBM
\paragraph{The \acrshort{ib} model for \acrshort{fsi}}
We consider a fully-coupled \acrshort{fsi} model based on the \acrshort{ib} method \cite{peskin1972flow, griffith2012immersed}. Dynamic balance between the fluid and solid model is enforced by defining the forcing term in the Navier Stokes' momentum balance \eqref{eq:ns-momentum} as:
\begin{equation}
    \forcing (\x, t) = \int_{\omegaszero} \divzero P \left (\X, t \right ) \dirac \left ( \x - \chi (\X, t)\right ) \mathrm{d} \X, \; \quad \text{with } \x \in \omegaf,  t \in (0, T),
    \label{eq:coupling-dynamic}
\end{equation}
and $\dirac$ is a 3D Dirac delta distribution. Conversely, continuity of velocity between the fluid and the structure is enforced through the following kinematic condition: 
\begin{equation}
    \pdv{\displ}{t} (\X, t) = \int_{\omegaf} \uf (\x, t) \dirac \left (\chi(\X, t) - \x \right ) \mathrm{d} \x  \; \quad \text{with } \X \in \omegaszero,  t \in (0, T).
    \label{eq:coupling-kinematic}
\end{equation}

% Numerical discretization
%fluid
\paragraph{Numerical methods}
We discretize the fluid dynamics problem in space with the \acrfull{fe} method with linear \acrshort{fe} spaces for both velocity and pressure. Time integration is carried out using an explicit third order Runge-Kutta method (with $\mathrm{CFL}=0.8$) and we impose the divergence-free constraint \eqref{eq:ns-continuity} with a fractional-step scheme. For additional details on the numerical schemes we used for the fluid problem, we refer the reader to \cite{lehmkuhl2019low}. The elastodynamic equation is solved with \acrshort{fe} method with quadratic \acrshort{fe} space, a generalized Newmark formulation as time advancing scheme.
To solve the fully-coupled \acrshort{fsi} problem, we use an explicit staggered coupling scheme: given the solution at a discrete time, we advance the time step by first solving the fluid dynamics problem, we interpolate $\uf$ to $\omegaszero$ (kinematic coupling condition \eqref{eq:coupling-kinematic}), we compute the displacement by integrating the interpolated velocity with a first-order Euler scheme, and we displace accordingly the solid domain; we compute the internal forces of the structure and we compute the forcing term \eqref{eq:coupling-dynamic}, enforcing the dynamic balance. For additional details on the \acrshort{ib}-\acrshort{fsi} scheme, we refer to \cite{oks2022fluid}. 

\paragraph{Computational setup}
The computational model is implemented in the multiphysics and multiscale \acrshort{fe} library \alya{} \cite{vazquez2016alya, santiago2018fully} developed at the Barcelona Supercomputing Center and designed to run efficiently on supercomputers in a \acrfull{hpc} framework to simulate tightly coupled problems. % \alya is one of the twelve codes of the \acrfull{ueabs}, thus complying with the higher standard in \acrshort{hpc}. 
We use a constant time step size equal to $\SI{5e-6}{\second}$ for both the fluid and solid problem. To execute efficiently the code, two \alya{} parallel instances are simultaneously executed, one for the fluid and one for the solid problem. At every time step, the two instances communicate with each other through a point-to-point scheme programmed using MPI-communicators. We use 568 and 256 cores for the fluid and solid problem, respectively. For additional details on the parallelization strategy of the \acrshort{av} \acrshort{fsi} model we deployed in \alya{}, we refer the reader to \cite{oks2022fluid}. 
We carry simulations on the ARCHER2 UK National Supercomputing Service\footnote{\url{https://www.archer2.ac.uk}} and the Nord 3 machine\footnote{\url{https://www.bsc.es/marenostrum}} at the Barcelona Supercomputing Center. We report computational times of the five simulations in \Cref{tab:cpu-time}: the average cost of simulating a heart cycle is around 10 hours. 

\begin{table}
	\centering
	\begin{tabular}{llccccc}
		\toprule
            \textbf{patient} & & P1 & P2 & P3 & P4 & P5 \\
            %\midrule
            \cmidrule(lr){3-7}
            $T$ (physical time) & [\si{\second}]& \num{10.6} & \num{10.2} & \num{7.1}& \num{7.1} & \num{6.8}  \\
            CPU time & [hours] & \num{129.6}  & \num{124.8} & \num{86.4} & \num{86.4} & \num{81.6}\\
		\bottomrule 
	\end{tabular}
	\caption{Computational times of the \acrshort{fsi} simulations. For each patient, we simulate ten heartbeats.  } 
		\label{tab:cpu-time}
\end{table}

\subsection{Feeding the FSI model with the \invivo{} flowrate}
\label{sec:methods-data-assimilation-flowrate}

\begin{figure}
    \centering
        \includegraphics[trim = {0cm 0cm 0cm 0cm}, clip, width=\textwidth]{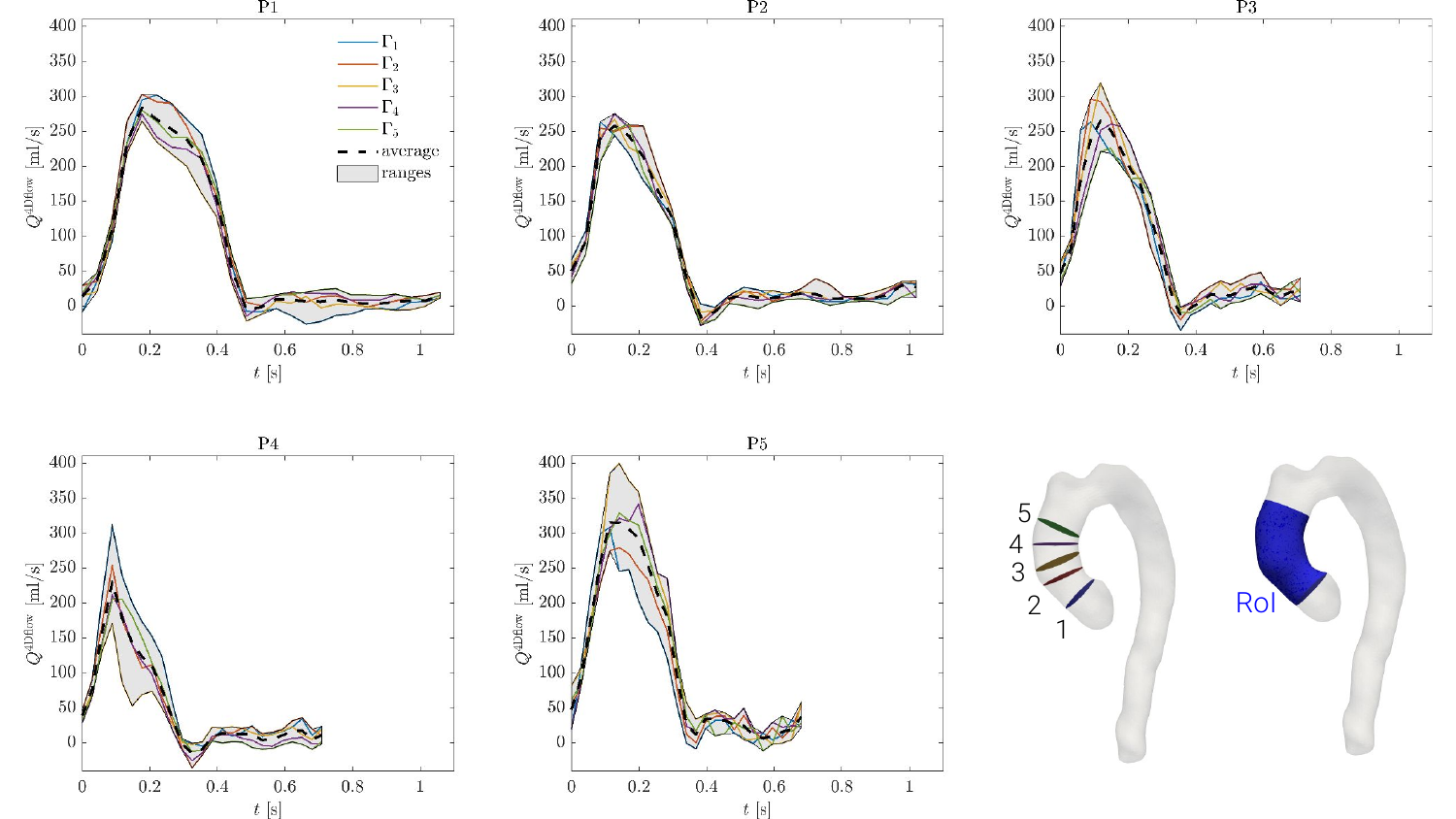}
    \caption{Flowrate in different planes computed from the \fourd{} data. In each plot, we report the flowrate computed in a different plane of the \acrshort{roi}, the average, and the ranges to highlight the spatial variability of these data.  }
    \label{fig:flowrate-4dflow}
\end{figure}
Considering that the paper aims to validate the blood flow in the region downstream the \acrshort{av}, we define a \acrfull{roi} as we display in \Cref{fig:flowrate-4dflow}. Several \acrshort{qois} are computed and averaged in this control volume.

To set boundary conditions on the inlet of the aorta, we compute the flowrate from the \fourd{} data. As displayed in \Cref{fig:flowrate-4dflow}, we compute the flowrate in five different planes $\Gamma_j$ belonging to the \acrshort{roi}  as
\begin{equation}
    \Qfourdgamma (t) = \int_{\Gamma_j} \langle \velfourd (\x, t) \rangle \cdot \bm n (\x, t)\, \mathrm{d} \x, \; \text{ with } j = 1, \dots, 5,
    \label{eq:Qdef}
\end{equation}
where $\bm n$ the normal to the plane $\Gamma_j$ considered. As illustrated in \Cref{fig:flowrate-4dflow}, notable variations exist in the computed flowrates among different sections, potentially attributed to the combined influences of aortic wall compliance and noise. To address these factors and include them in the computational model, we compute the average of these flowrates and use it as our inlet flowrate: 
\begin{equation}
    \Qin (t) = \Qfourdavg(t) = \frac{1}{5}\sum_{j=1}^{5} \Qfourdgamma (t).
     \label{eq:Qavg} 
\end{equation}
In \Cref{fig:flowrate-4dflow}, we show shaded regions to accommodate the flowrate's variability. Moreover, to ensure that the \acrshort{av} remains closed during diastole, we set $\Qin=0$ during this phase.  Furthermore, since the \fourd{} flowrate is characterized by a poorer time resolution compared to the one required by the \acrshort{fsi} problem, we use B-splines to approximate and smooth this temporal function in time. 

\subsection{Calibration of the Young modulus of the \acrshort{av}}
\label{sec:results-calibration-young}

\begin{figure}[t]
    \centering
    \includegraphics[trim = {0cm 7cm 0cm 7cm}, clip, width=\textwidth] {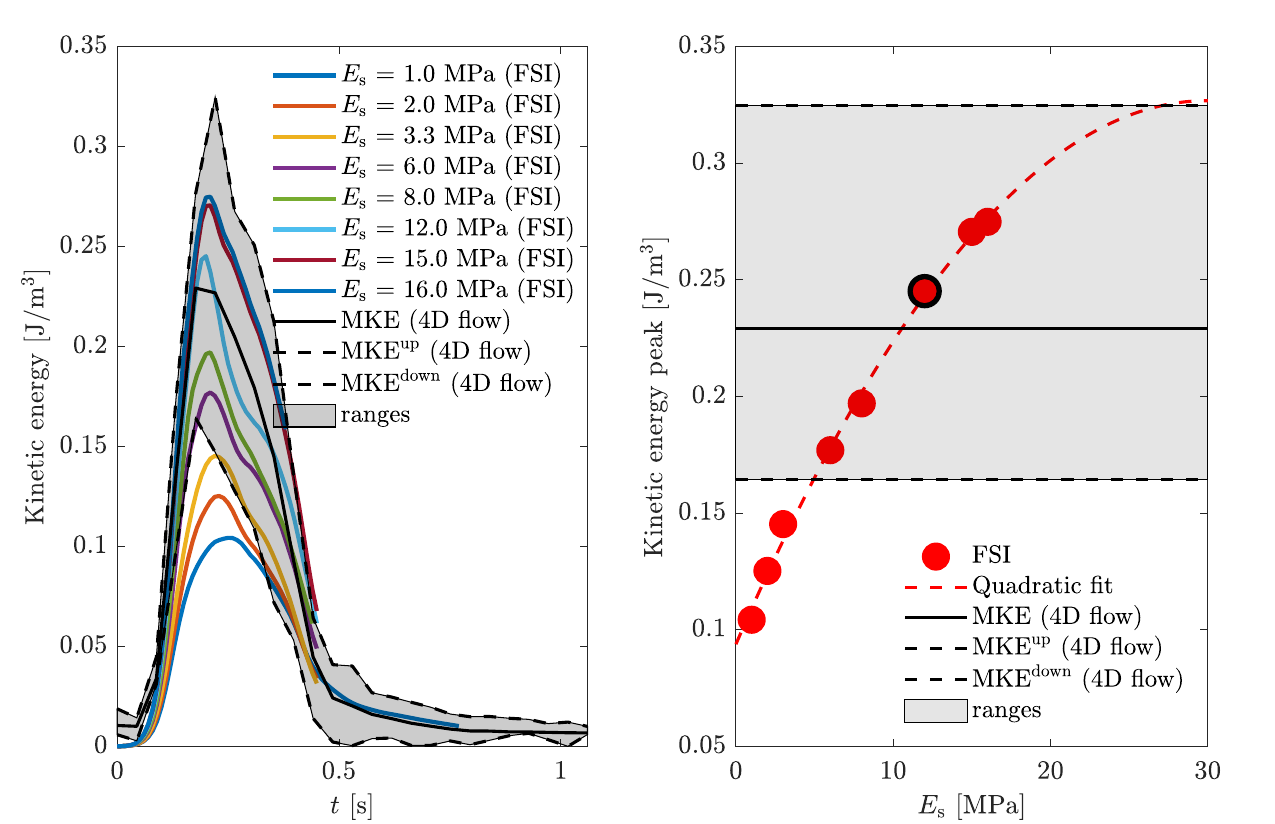}
    \caption{Calibration of the Young modulus $\young$ for P1 to get an accurate representation of the kinetic energy in the \acrshort{roi}. Left: temporal evolution of the \acrshort{mke} obtained by eight different Young modulus and the \invivo{} \fourd{} data. Right: peak of kinetic energies for the eight setups of the FSI model against peak from \fourd{} data (with corresponding range), with quadratic fit of the FSI results. }
    \label{fig:tuning-young-modulus}
\end{figure}
\begin{figure}[t]
    \centering
    \includegraphics[trim = {0cm 1.5cm 0cm 0cm}, clip, width=\textwidth] {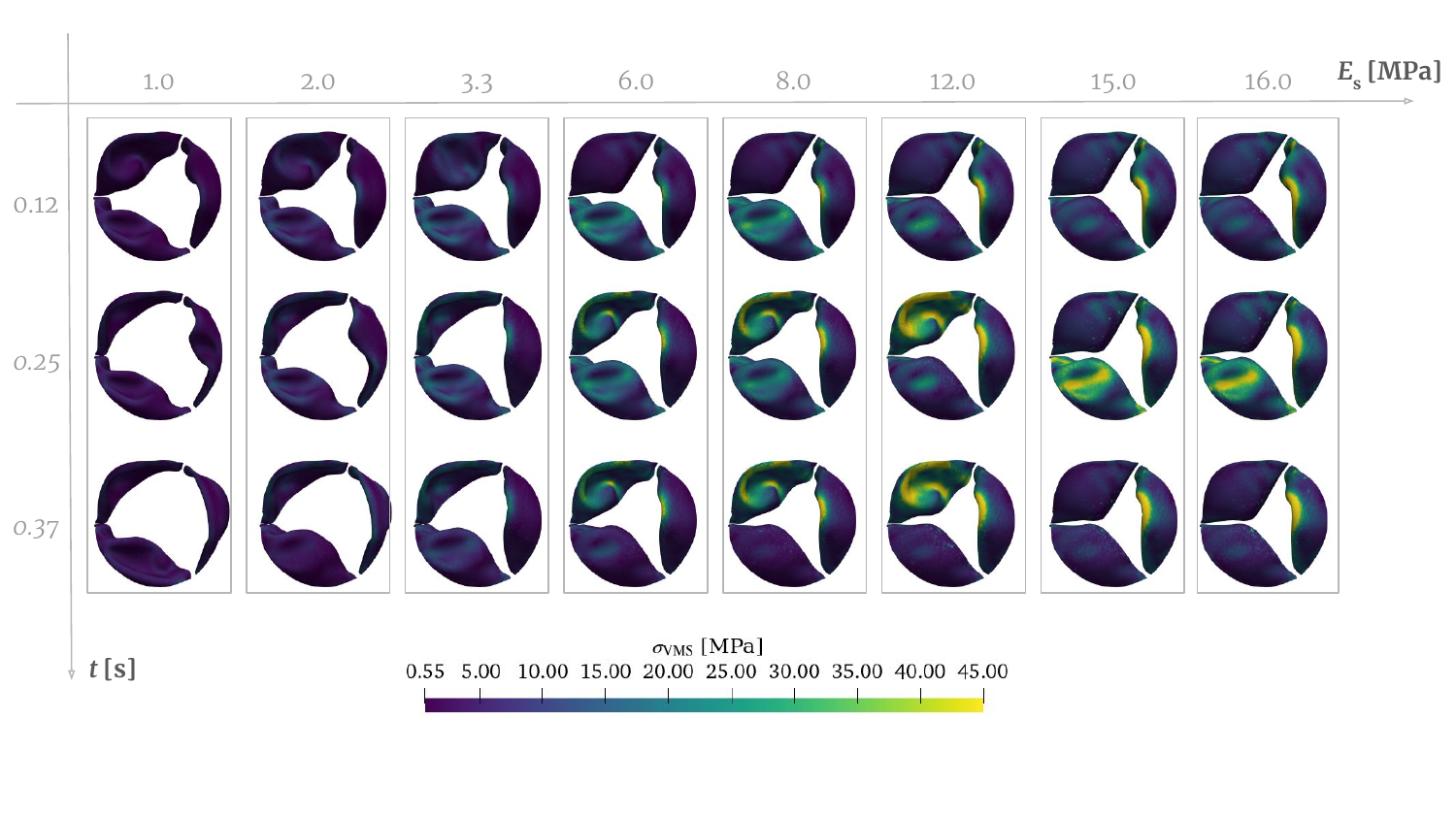}
    \caption{\acrshort{av} in three different instants of the heart cycle ($t=\SI{0.12}{\second}$, $t=\SI{0.25}{\second}$, and $t=\SI{0.37}{\second}$) for different values of the Young modulus $\young$. The \acrshort{av} is colored according to the von Mises stress. }
    \label{fig:tuning-young-modulus-3dvalve}
\end{figure}

To carry out \acrshort{fsi} simulations of the \acrshort{av}, it is crucial to accurately select the parameters of the structural model of the valve, parameters that might be patient-specific and also related to the stenosis level. We set the density value to $\rhos=\SI{1000}{\kilo\gram\per\meter\cubed}$. Regarding the Lamé constants -- which are related to the Young modulus $\young$ and the Poisson's ratio $\poisson$ -- we fix $\poisson = 0.3$, and we calibrate $\young$ to match the \acrfull{mke} per unit mass of the \fourd{} data downstream of the \acrshort{av} within the \acrshort{roi} for patient P1. 

\paragraph{Spreading the uncertainty of the flowrate onto the \acrshort{mke}}

To spread the uncertainty of the inlet flowrate into the kinetic energy (see \Cref{fig:flowrate-4dflow}), in the following, we establish ranges of uncertainty for the \acrshort{mke}. Recalling that  $\rhof$ is constant, we define the \acrshort{mke} per unit mass as: 
\begin{equation}
    \mke(t) = \frac{1}{2|\Omegaroi|} \int_{\Omegaroi} |\langle \velfourd (\x, t) \rangle |^2 \,  \dx,
    \label{eq:mke}
\end{equation}
where $|\Omegaroi|$ is the volume of the \acrshort{roi}.
We then define the upper and lower \acrshort{mke} as
\begin{align}
    \mkeup(t) & = \mke(t) + \deltaup(t) \mke(t), \\
    \mkedown(t) & = \mke(t) - \deltadown(t) \mke(t),
\end{align}
where $\deltaup \in [0, 1]$ and $\deltadown\in [0, 1]$ are the upper and lower relative variations with respect to the averaged \acrshort{mke}. 
These ranges encompass variations in aortic blood flow rates arising from \fourd{} acquisitions noise and aortic compliance, which is not considered in the computational model. To compute kinetic energy uncertainty, we derive minimum, maximum, and average velocities from flow measurements, and then integrate the square of those data into the kinetic energy computation. Further elaboration on this process follows. Considering \Cref{eq:Qdef}, we define the minimum and maximum measured flowrates for all the $N$ time frames of the \fourd{} acquisition as:
\begin{equation}
    \Qfourdmin(t_i) = \min_j \Qfourdgamma(t_i), \; \qquad
    \Qfourdmax(t_i) = \max_j \Qfourdgamma(t_i), \; \qquad 
    \text{for all } i = 1, \dots, N.  \label{eq:QminQmax} 
\end{equation}

Let $\Amin(t_i)$ and $\Amax(t_i)$ be the area of the section corresponding to the minimum and maximum flowrate computed in \Cref{eq:QminQmax}, and let  $\Aavg(t_i)$ be the average area of the sections at time $t_i$, with $i = 1, \dots, N$. We define the average, minimum, and maximum velocity -- constant in space -- corresponding to the flowrates in Equations \eqref{eq:Qavg} and \eqref{eq:QminQmax} as:
\begin{equation*}
\uavg(t_i) = \frac{\Qfourdavg(t_i)}{\Aavg(t_i)}, \; \quad
\umin(t_i) = \frac{\Qfourdmin(t_i)}{\Amin(t_i)}, \; \quad
\umax(t_i) = \frac{\Qfourdmax(t_i)}{\Amax(t_i)}, \quad \text{for all } i = 1, \dots, N.
\end{equation*}

To compute the corresponding \acrshort{mke}s in the \acrshort{roi}, since $\uavg$, $\umin$, and $\umax$ are constant in space, the associated kinetic energies are simply defined as:
\begin{equation*}
    \mkeavgtilde(t_i) = \frac{1}{2|\Omegaroi|}  \, \uavg^2(t_i), \; \quad
    \mkemintilde(t_i) = \frac{1}{2|\Omegaroi|}  \, \umin^2(t_i), \; \quad
    \mkemaxtilde(t_i) = \frac{1}{2|\Omegaroi|}  \, \umax^2(t_i). 
\end{equation*}
We then define the relative upper and lower variations as:
\begin{align*}
    \deltaup(t_i) & = \frac{\left |\mkemaxtilde(t_i) - \mkeavgtilde(t_i) \right |}{\max\left (\mkemaxtilde(t_i), \mkeavgtilde(t_i) \right )}, \; \text{for all } i = 1, \dots, N,
    \\
    \deltadown(t_i) & = \frac{\left | \mkemintilde(t_i) - \mkeavgtilde(t_i) \right |}{\max \left (\mkemintilde(t_i), \mkeavgtilde(t_i) \right )}, \; \text{for all } i = 1, \dots, N.
\end{align*}
Notice that we define the relative changes as above to avoid division by very small numbers \cite{tornqvist1985should}.

\paragraph{Calibration of the Young modulus}

We calibrate the Young modulus for patient P1 using the following values: 
\[
\young = \left [  1.0, \, 2.0, \, 3.3, \, 6.0, \, 8.0, \, 12.0, \,  15.0, \,  16.0 \right ] \si{\mega\pascal}. 
\]
For this calibration process, simulations were conducted for less than one heartbeat since the tuning of the Young modulus is based solely on the peak kinetic energy. Thus, to be precise, we compare here the \acrshort{mke} (i.e. with phase-averaged velocity) from \fourd{} against the kinetic energy (i.e. from the velocity in the first beat) from \acrshort{fsi}. 

\Cref{fig:tuning-young-modulus} illustrates the results of our calibration process: as the Young modulus increases, the computed values approach our reference values from \fourd{} data. It is important to note that, for the \fourd{} data, we present both the actual \acrshort{mke} computed in the \acrshort{roi} and the corresponding variations due to the uncertainty of the flowrate, as explained earlier. In \Cref{fig:tuning-young-modulus-3dvalve}, we display the \acrshort{av} in three different instants of the heartbeat for the values considered of the Young modulus. As expected, as $\young$ increases, we observe a smaller orifice area, resulting in a faster blood jet (since the inlet flowrate remains constant). Furthermore, as the Young modulus increases, we compute larger values of the von Mises stress $\sigma_\mathrm{VMS}$, an indicator that can be used to assess the calcification level of the \acrshort{av} \cite{oks2022fluid}. 

The calibration study allows us to establish that $\young = \SI{12}{\mega\pascal}$ accurately replicates the  \fourd{} data for P1, providing computational kinetic energy values within the range of variation of the \invivo{} \fourd{} and closer to the computed MKE. 
Consequently, we select this value of the Young modulus for the entire study. Consistency is maintained by applying the same Young modulus value to the remaining patients, given their similar severity levels of \acrshort{as}. Moreover, this decision expedites the calibration process across the entire study.

\section{Results}
\label{sec:results}
We present the results of our study. We validate our results against \invivo{} data in \Cref{sec:results-validation}, whereas in \Cref{sec:results-additional}, we present additional \acrshort{qois} computed with the \acrshort{fsi} simulations useful for \acrshort{as} assessment.

\subsection{Validation}
\label{sec:results-validation}

\begin{figure}
    \centering
    \includegraphics[trim = {0cm 9.5cm 0cm 0cm}, clip, width=\textwidth]{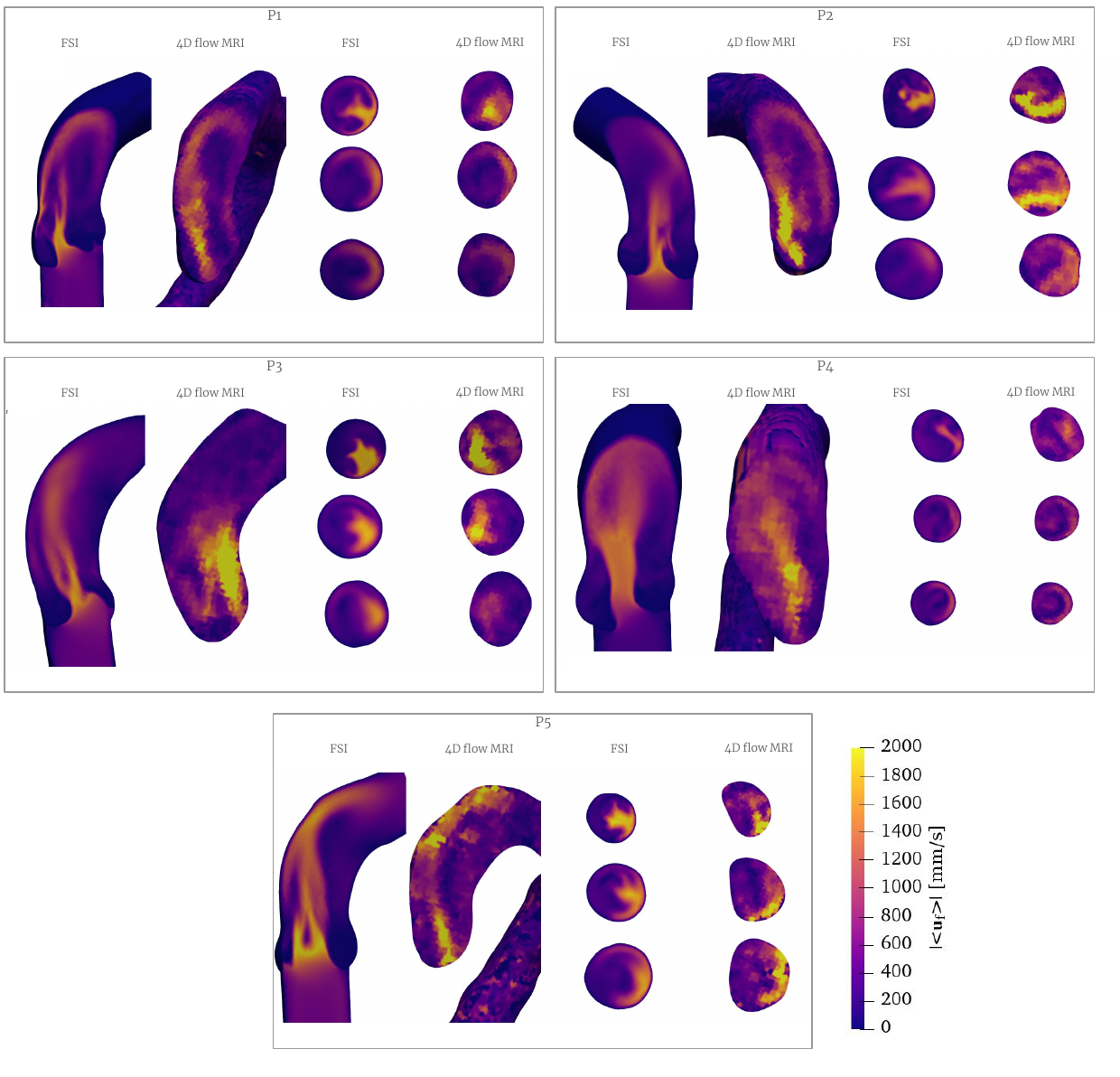}
    \caption{Validation of phase-averaged velocity magnitude against \fourd{} data at the systolic peak for the five patients. For each patient, results are compared on a clip in the longitudinal direction of the ascending aorta, and on three slices in the \acrshort{roi}.}
    \label{fig:validation-velocity-clip-slices}
\end{figure}

% velocity
We validate the \insilico{} results against \fourd{} velocity data. In particular, since the \invivo{} data consists on phase-averaged velocity fields, we run our \acrshort{fsi} simulations for ten heartbeats, we discard the first two cycles to remove the influence of null initial conditions, and we phase-average the velocity in the remaining eight cycles as \cite{zingaro2021hemodynamics}:
\begin{equation*}
    \langle \uf (\x, t) \rangle = \frac{1}{8} \sum_{n = 3}^{10} \uf (\x, t+(n-1)\Thb).
\end{equation*}
\Cref{fig:validation-velocity-clip-slices} illustrates the validation of $\uphase$ for all five patients at the systolic peak. Specifically, we compare the velocity along a clip in the longitudinal direction of the aorta and three slices within the \acrshort{roi}. Our findings indicate that, except for P2 and P3, the FSI model accurately reproduces the flow patterns observed in the \invivo{} data. This includes a faithful capture of the strong jet crossing the \acrshort{av}, along with its impingement on the wall. Additionally, when considering the slices within the \acrshort{roi}, we observe a similar spatial velocity distribution between the computational results and the \invivo{} data. We believe that the discrepancy in the velocity patterns observed for P2 and P3 might be due to the use of the same valve geometry among different patients. 

% mke
\begin{figure}
    \centering
    \includegraphics[trim = {0cm 6.5cm 0cm 6.5cm}, clip, width=\textwidth]{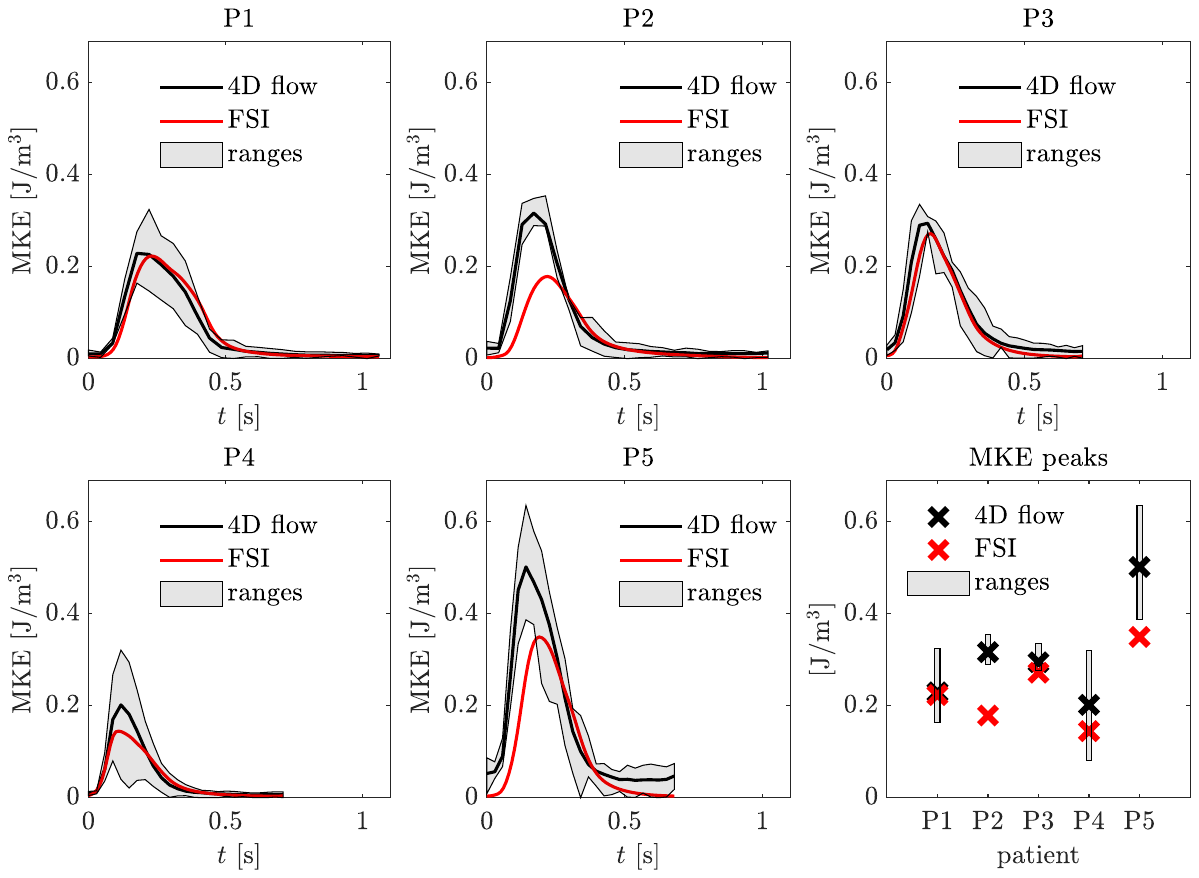}
    \caption{Validation of \acrshort{mke} against \fourd{} data for the five patients.}
    \label{fig:validation-mke}
\end{figure}
To quantitatively assess the disparity between the computational results and the 4D flow MRI data, we calculate the \acrshort{mke} within the region of interest using \Cref{eq:mke} with $\uphase$. \Cref{fig:validation-mke} presents a comparison of the \acrshort{mke}, incorporating the ranges of variations detailed in \Cref{sec:methods-data-assimilation-flowrate}.As expected, larger variations of the flowrate correspond to larger variations of the \acrshort{mke}. Notably, with the exception of P2, the computational results exhibit a temporal evolution closely resembling that of the \invivo{} data, with most patients' computational \acrshort{mke} falling within the range of variations observed in the 4D flow MRI data. The discrepancy for patient P2 is consistent with the larger velocity observed in \Cref{fig:validation-velocity-clip-slices}. Interestingly, despite discrepancies in the flow patterns of P3, the model accurately reproduces its kinetic energy. In \Cref{fig:validation-mke}, we also provide a comparison of the MKE peaks for the five patients, showing that the peaks are always within the ranges of uncertainty, except for P2 and P5: for both patients, we underestimate the \invivo{} MKE peak. Notice also that, since the calibration process is carried out for P1 only, and the same value of $\young$ is then used for the remaining patients, it is reasonable to expect a better accuracy for P1 than the remaining ones. 

% ava
\begin{figure}
    \centering
    \includegraphics[trim = {0cm 8.5cm 0cm 7.5cm}, clip, width=\textwidth]{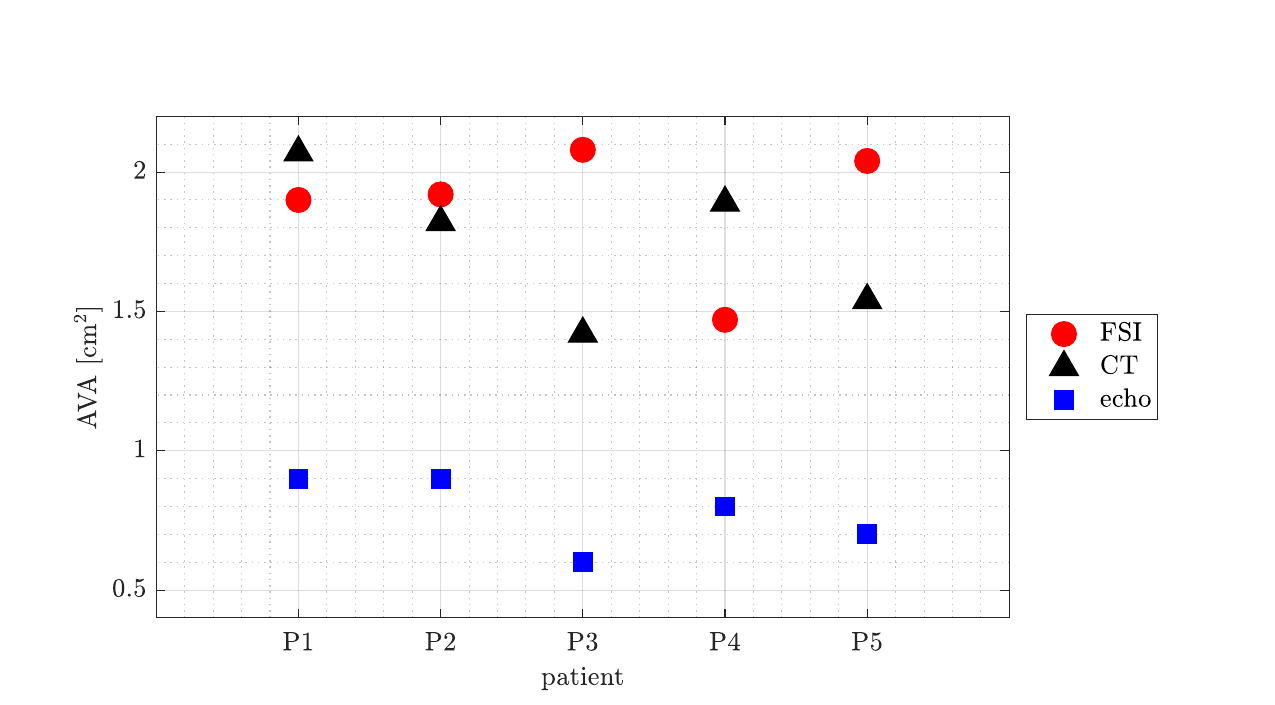}
    \caption{\acrshort{ava} measurements: comparison between \acrshort{fsi}, \acrshort{ct} measurement and echo measurement.}
    \label{fig:ava}
\end{figure}

We compare the orifice area (or \acrshort{ava}) using computational and \invivo{} measurements. From the \acrshort{fsi} simulations, we compute the \acrfull{goa} as the area enclosed by the projection of the leaflet commissure curves on the aortic cross section \cite{oks2022fluid}. Furthermore, clinicians carried out echocardiography measurements to compute the \acrshort{ava} via continuity equation method \cite{bonow1998acc}. In \Cref{fig:ava}, we compare the \acrshort{ava} using \acrshort{fsi} and echo measurements: \insilico{} simulations give \acrshort{ava} results in the order of two/three times larger than \invivo{} measurements. Differently, by measuring directly the \acrshort{ava} from the static \acrshort{ct} images, we obtain values that are much more in line with the computational results, resulting in a mean relative error of 23\%. This explains that the main source of error might come from the measurement method (continuity-based vs. geometrical).

% sss and wss
\begin{figure}
    \centering
    \includegraphics[trim = {0cm 9.0cm 0cm 5.5cm}, clip, width=\textwidth]{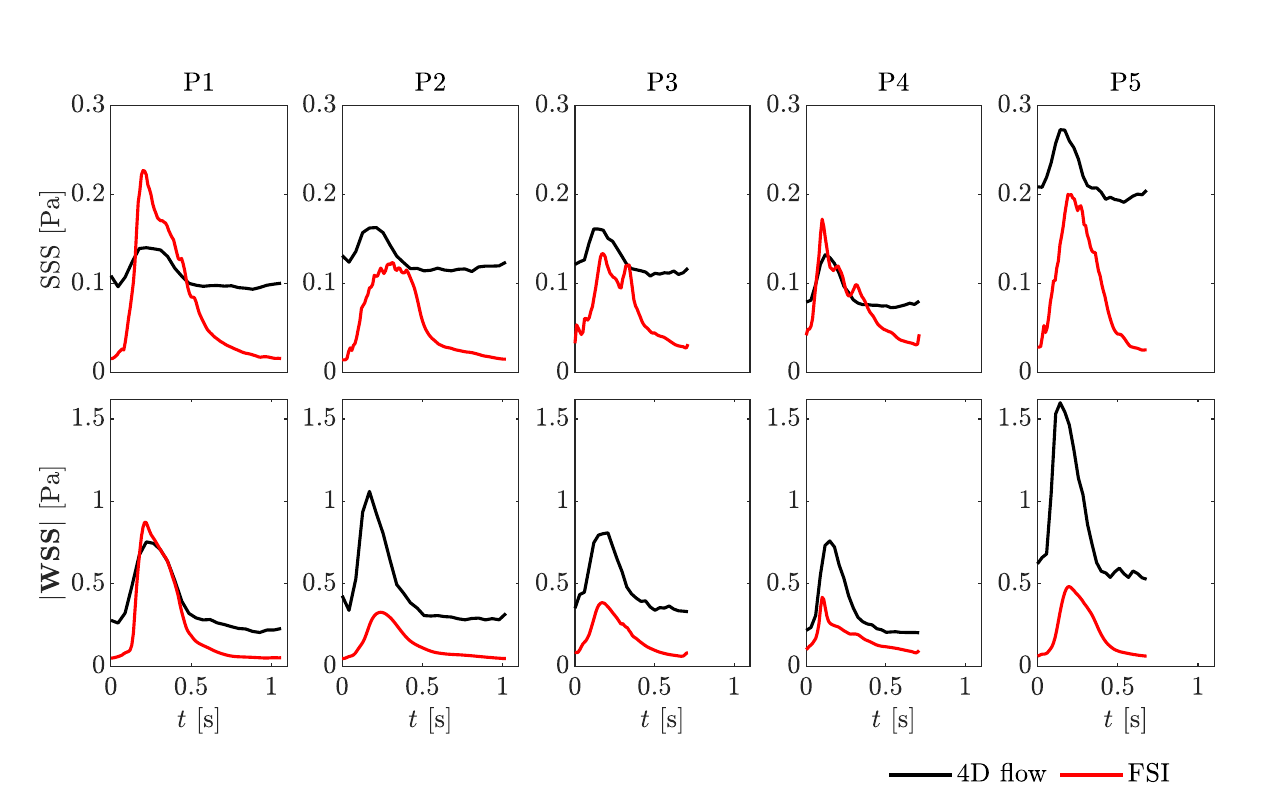}
    \caption{Validation of space-averaged \acrshort{sss} (top) and space-averaged \acrshort{wss} magnitude (bottom) against \fourd{} data for the five patients.}
    \label{fig:validation-sss-wss}
\end{figure}

\begin{figure}
    \centering
    \includegraphics[trim = {0cm 2.5cm 0cm 0cm}, clip, width=\textwidth]{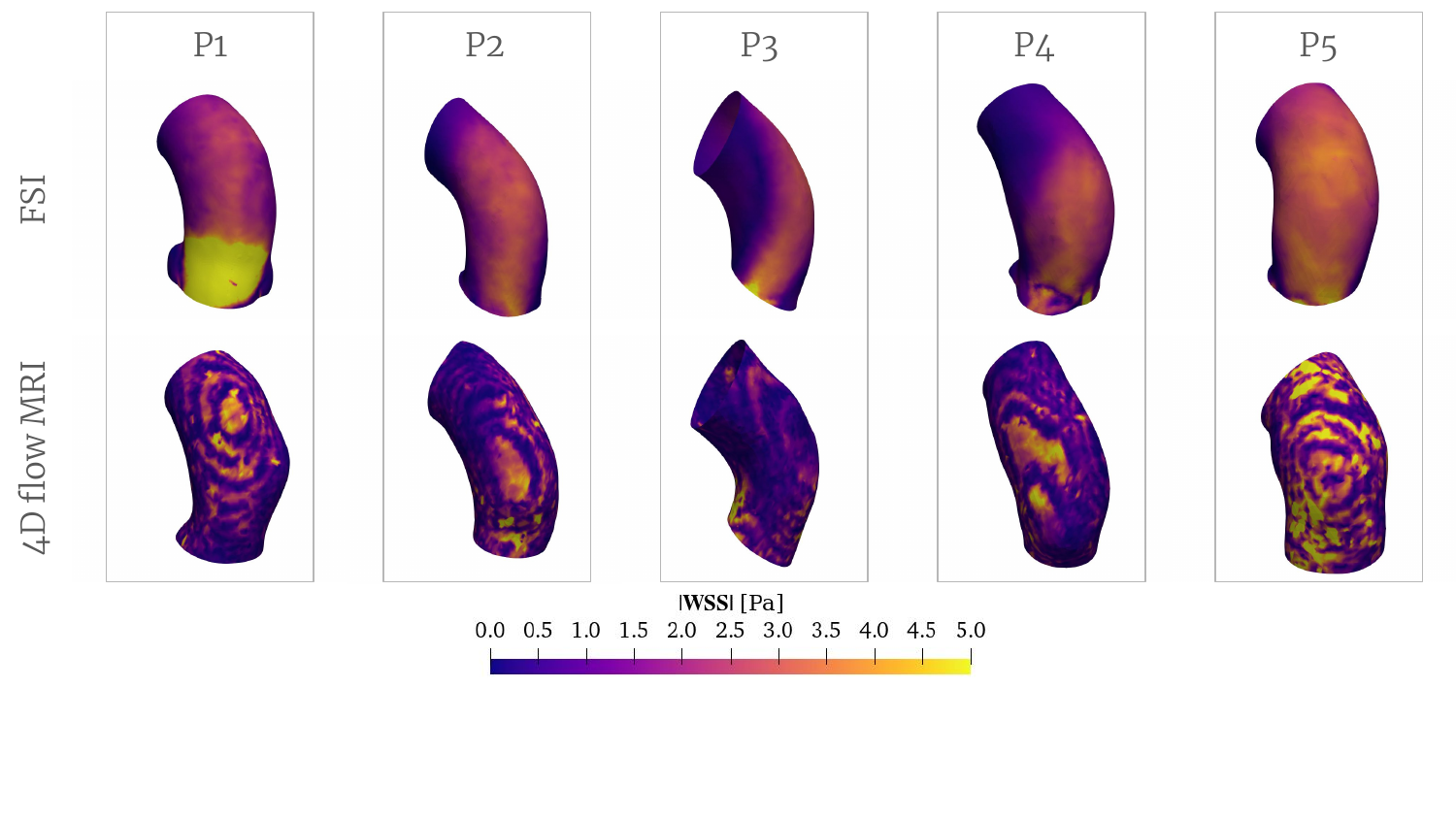}
    \caption{Validation of \acrshort{wss} in the \acrshort{roi}. Top: \acrshort{fsi}; bottom: \fourd{}.}
    \label{fig:validation-wss}
\end{figure}

To assess the reliability of the computational model in terms of shear stress, we compute several quantities inside the fluid domain and on its boundary.  Let $\tauf(\uphase) = \muf (\grad \uphase + \grad^T \uphase)$ be the viscous stress tensor, we compute the \acrfull{sss} as \cite{zhang2020computational}:
\begin{equation*}
    \sss (\x, t) = \sqrt{\frac{1}{6}\sum_{i=1}^{3}\sum_{\substack{j=1 \\ j \neq i}}^3 \left ( \left ( \tauf_{ii} - \tauf_{jj}\right )^2 + \tauf_{ij}^2 \right ) } , \quad \text{ in } \omegaf \times (0, \Thb)
\end{equation*}
and the \acrfull{wss} on the boundary of the fluid domain as:
\begin{equation*}
    \wss (\x, t) = \tauf (\uphase) \nf - \left ( \tauf (\uphase) \nf \cdot \nf \right ) \nf, \quad \text{ on } \partial \omegaf \times (0, \Thb).
\end{equation*}

\begin{figure}
    \centering
    \includegraphics[trim = {0cm 0cm 0cm 0cm}, clip, width=\textwidth]{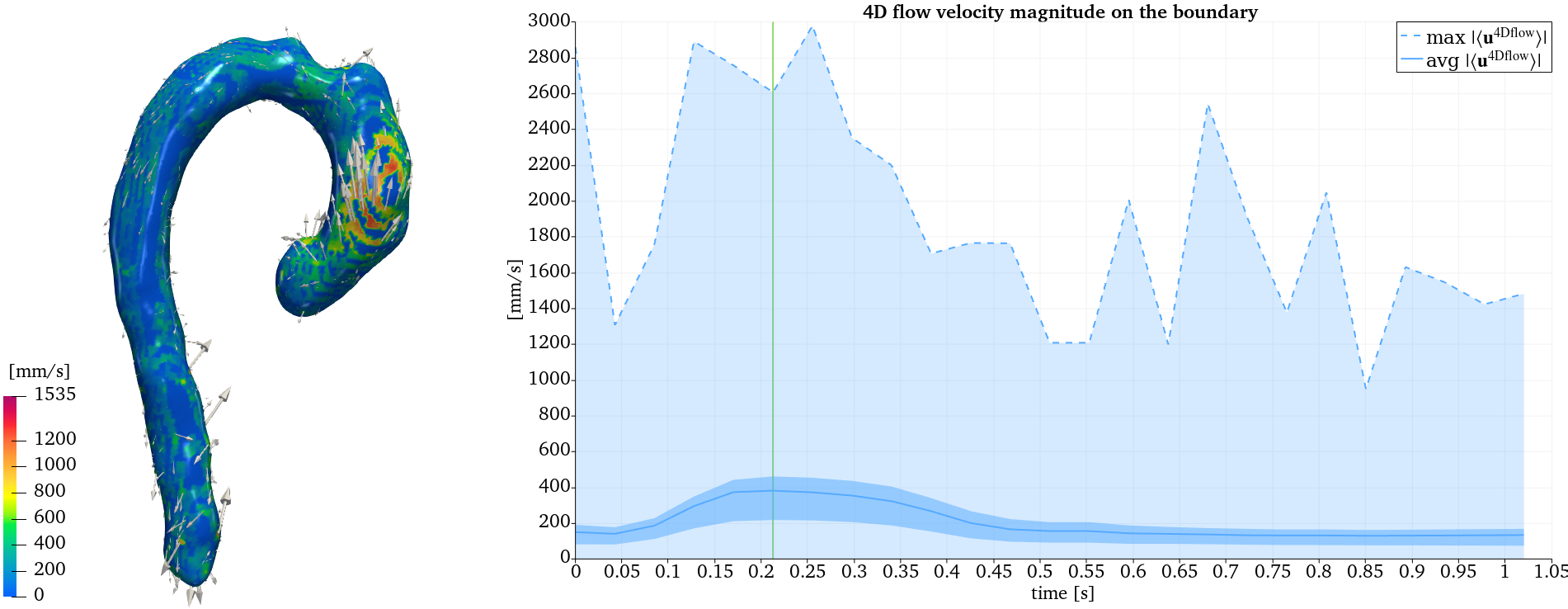}
    \caption{\fourd{} velocity on the boundary for P1. On the left, a visualization of the velocity on the wall of the aorta at the systolic peak. To the right, the time series of the maximum, average and quartile velocity on the aorta boundary. }
    \label{fig:velocity-boundary-P1}
\end{figure}

We compare the \acrshort{sss} and the \acrshort{wss} for the five patients against \fourd{} data: in \Cref{fig:validation-sss-wss} we compare the space-averaged \acrshort{sss} and the space-averaged \acrshort{wss} magnitude. In all the cases, the \acrshort{fsi} simulation produces profiles similar to the \fourd{}, where large values of both indicators are measured during the ejection phase. However, some discrepancies in terms of values are observed. For \acrshort{sss}, there are cases where the \acrshort{fsi} underestimates the \fourd{} data, while in others, the computational results yield larger values. Concerning \acrshort{wss}, a general trend emerges where, except for P1, the FSI produces lower values.  This trend aligns with the observations for the \acrshort{mke} in \Cref{fig:validation-mke}, potentially also accentuated by the lower resolution of the \fourd{} data compared to the \acrshort{fsi} simulations, which, along with the noise characterizing the \invivo{} data, may highly affect the accuracy of velocity gradient computations. As we illustrate in \Cref{fig:velocity-boundary-P1}, the \fourd{} data are often characterized by large and unrealistic velocities and velocity gradients at the wall in localized regions. Thus, this strongly affects the reliability of \invivo{} stress quantities at the wall, especially since velocity derivatives are involved. 

In \Cref{fig:validation-wss}, we report the \acrshort{wss} at the systolic peak on the boundary of the \acrshort{roi}. The field computed with the \fourd{} data appears not very regular. Nonetheless, generally speaking we can observe that, for both \acrshort{fsi} and \fourd{}, as the blood accelerates through the \acrshort{av} leaflets and impinges the aortic wall, larger values of the \acrshort{wss} are measured. In P1, we observe larger values of the stress in the lower part of the \acrshort{roi} that are not found in the \invivo{} data. Furthermore, for P3, we fail to reproduce the region with larger \acrshort{wss} and this is consistent with the discrepancy of the flow pattern observed in  \Cref{fig:validation-velocity-clip-slices}.

% vorticity, enstrophy
\begin{figure}
    \centering
    \begin{subfigure}{\textwidth}
            \includegraphics[trim = {0cm 3.0cm 0cm 0cm}, clip, width=\textwidth]{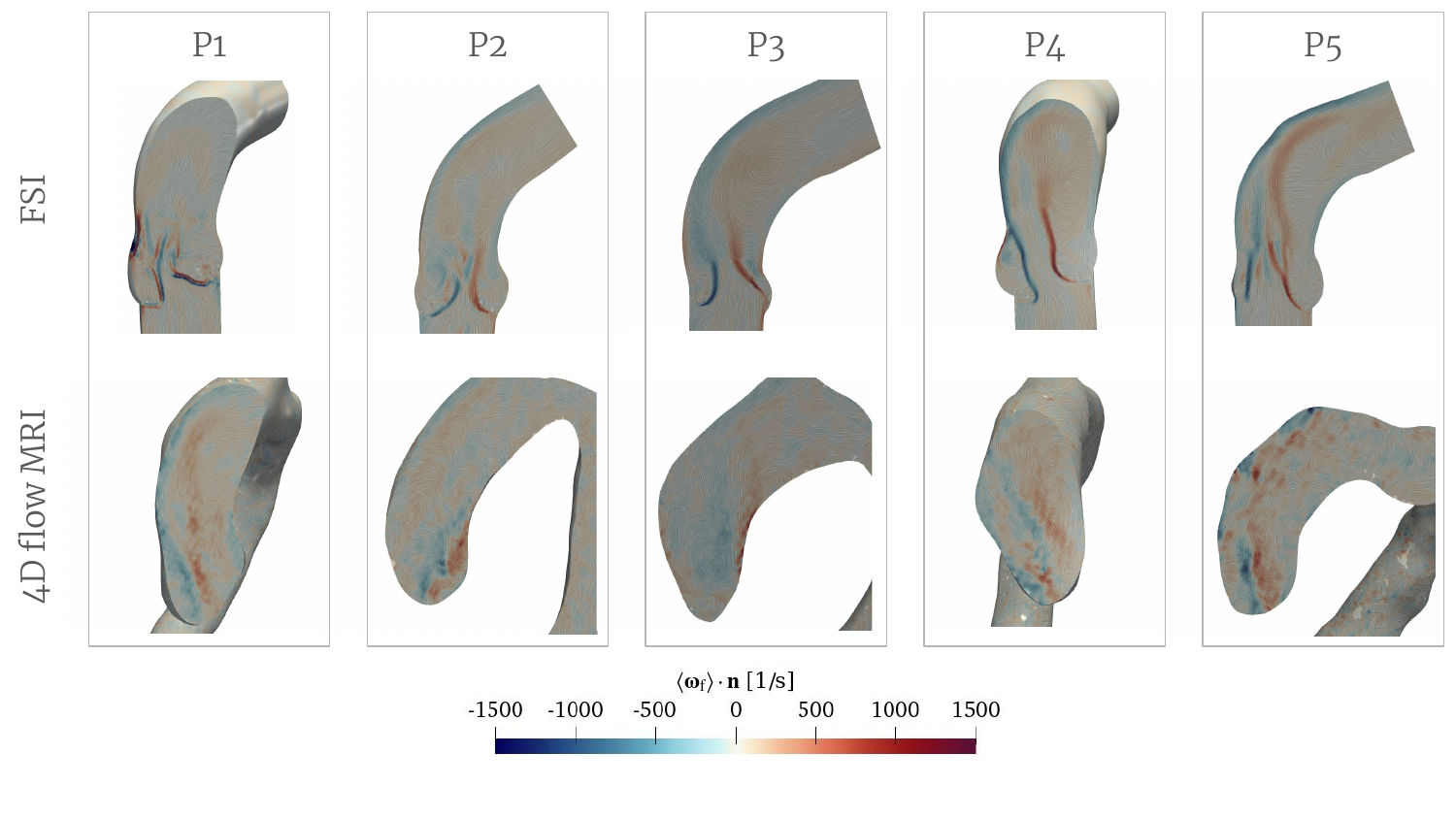}
    \end{subfigure}
    \\
    $\vorticity \cdot \bm n \, [\si{\per \second}]$
    \vspace{0.2cm}
    \begin{subfigure}{\textwidth}
            \includegraphics[trim = {0cm 1.0cm 0cm 12.2cm}, clip, width=\textwidth]{figures/validation-vorticity.pdf}
            \caption{}
            \label{fig:validation-vorticity-normal}
    \end{subfigure}
    \\
    \begin{subfigure}{\textwidth}
        \includegraphics[trim = {0cm 10cm 0cm 10cm}, clip, width=\textwidth]{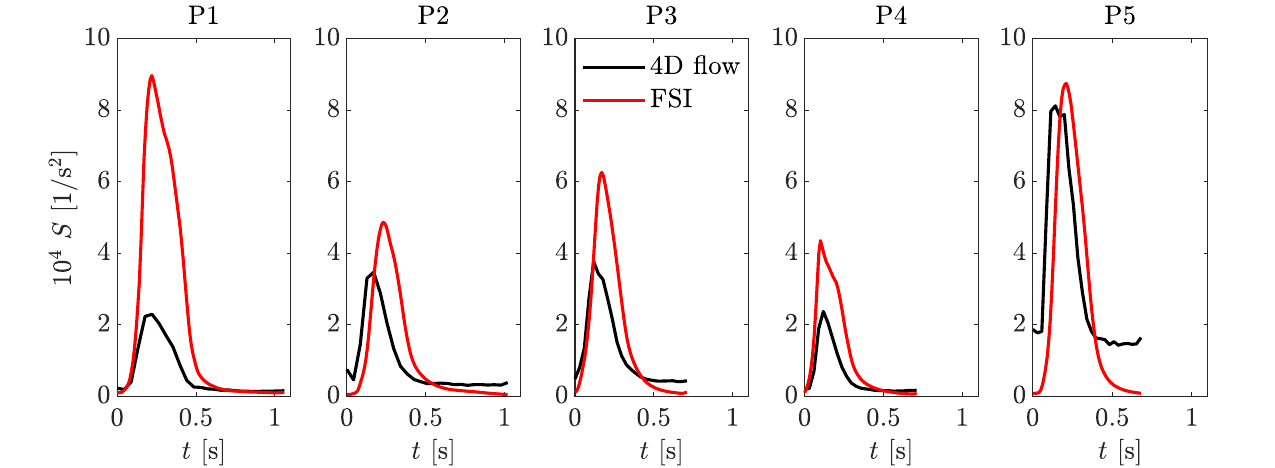}
            \caption{}
            \label{fig:validation-S-H}
    \end{subfigure}
    \caption{Validation of vorticity and enstrophy a) validation of the projection of the vorticity in the normal direction of a plane of the aorta (top, \acrshort{fsi}; bottom, \fourd{}). (b) Validation of enstrophy $S$: in red, \acrshort{fsi}; in black, \fourd{}.}
    \label{fig:validation-vorticity-normal-S-H}
\end{figure}

Next, we assess the accuracy of the model in terms of its ability to capture the vorticity and the enstrophy. Let $\vorticity = \grad \times \uphase$ be the vorticity computed with the phase-averaged velocity; we compute the enstrophy per unit mass in the \acrshort{roi} as:
\begin{equation*}
    S(t) = \frac{1}{|\Omegaroi|}\int_{\Omegaroi} |\vorticity|^2 \dx.
\end{equation*}

In \Cref{fig:validation-vorticity-normal}, we display the projection of the vorticity in the normal direction of a plane of the aorta ($\vorticity \cdot \bm n$) using the \acrshort{fsi} simulations and \fourd{} data. In the same plane, we also report surface \acrfull{lic} of the phase-average velocity, to better appreciate the formation of vortices. In all cases, we can observe shear layers forming on the leaflets of the \acrshort{av} producing counter-rotating vortices into the Valsalva sinuses, a well-known physiological pattern \cite{keele1952leonardo, robicsek1991leonardo}. Similar patterns between \acrshort{fsi} and \fourd{} are observed for all the patients except P3, consistently with the previous results on velocity and \acrshort{wss}. Furthermore, due to the lower resolution and the noise of the \invivo{} data, the numerical simulations provide a much more detailed visualization of the vorticity field, allowing for a better inspection of vortices and shear layers. 

In \Cref{fig:validation-S-H}, we present the space-averaged enstrophy. Across all cases, our computational model accurately tracks the temporal evolution observed in the \invivo{} data. However, we note that the enstrophy values derived from the \acrshort{fsi} simulations consistently surpass those from the \invivo{} measurements. This discrepancy may stem from the lower reliability of \fourd{} in computing velocity derivatives, as we discussed for the \acrshort{wss} validation.

\subsection{Additional \acrshort{qois} from \acrshort{fsi} simulations}
\label{sec:results-additional}

After validating the results, we demonstrate the utility of \acrshort{fsi} in computing additional \acrshort{qois} essential for assessing \acrshort{as}. While \fourd{} enables the quantification of several phase-averaged hemodynamic parameters related to blood flow, including mean velocity, \acrshort{mke}, and shear stress resulting from mean flow, the following section focuses on extracting further \acrshort{qois} from \acrshort{fsi} simulations specifically beneficial for \acrshort{as} assessment, beyond the scope of validation.

\acrshort{as} patients are often characterized by increased turbulence that can be estimated by the \acrfull{tke} of the blood flow \cite{toggweiler2022turbulent}. By defining the root mean squared velocity as  $\bm u ' (\x, t) = \sqrt{\langle \uf ^2 (\x, t)\rangle - \langle \uf(\x, t) \rangle^2}$, we compute the \acrshort{tke} as \cite{chnafa2014image, zingaro2021hemodynamics} 
\begin{equation*}
    \mathrm{TKE}(t) = \frac{1}{2 |\Omegaroi|} \int_{\Omegaroi}  \bm u '(\x, t)^2 \dx.
\end{equation*}

\begin{figure}[t]
    \centering
    \begin{subfigure}{\textwidth}
        \includegraphics[trim = {0cm 10.5cm 0cm 5.5cm}, clip, width=\textwidth]{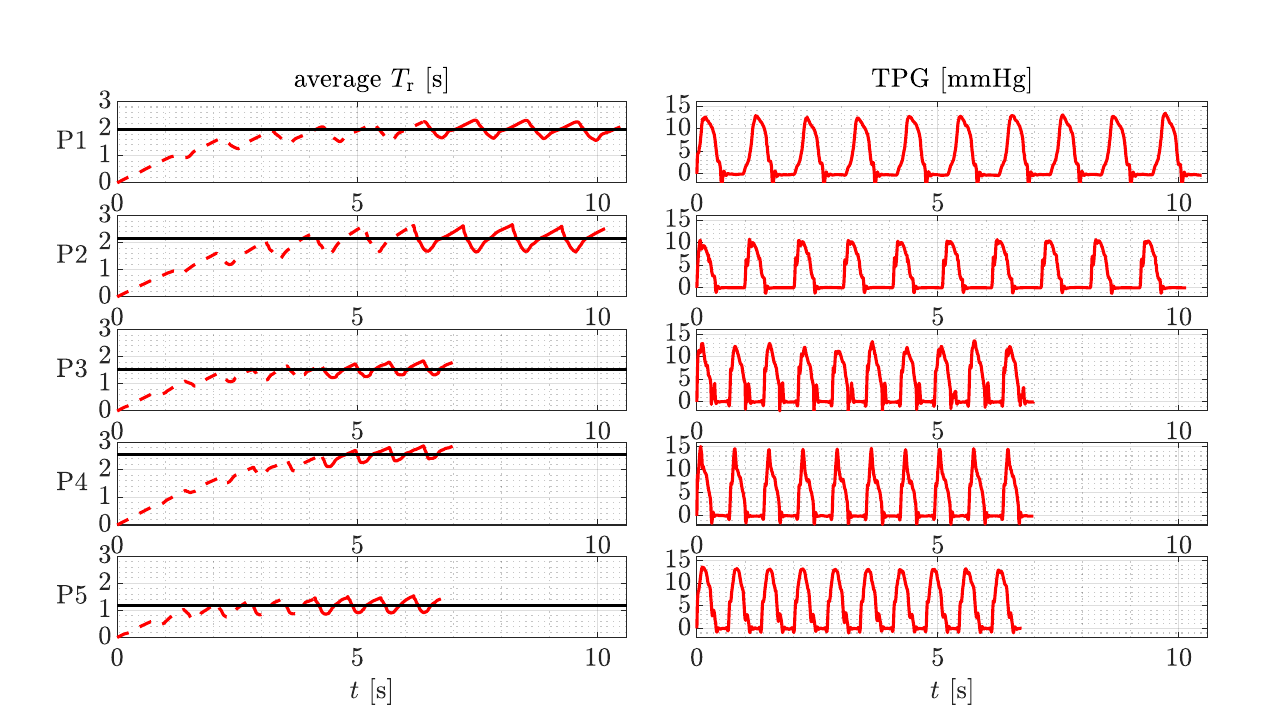}
    \end{subfigure}
    \begin{subfigure}{\textwidth}
        \includegraphics[trim = {0cm 6.5cm 0cm 16.0cm}, clip, width=\textwidth]{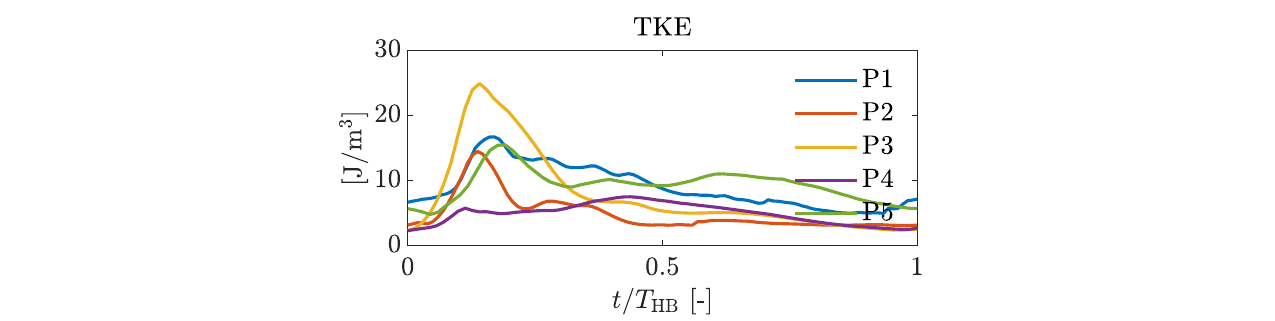}
    \end{subfigure}
    \caption{\acrshort{qois} from \acrshort{fsi} simulations for the five patients: on the left, average $\Tres$ for ten beats, in black, we report the average $\Tres$ by averaging the solution in the last four beats (solid red line) discarding the first six beats (dashed red line); on the right, the \acrshort{tpg} for ten beats; on the bottom, the \acrshort{tke} on a representative cycle. }
    \label{fig:rt-tpg-tke}
\end{figure}

Furthermore, since patients with \acrshort{as} are characterized by large pressure gradients \cite{otto2021acc, baumgartner20172017}, we compute the \acrshort{tpg} by averaging the pressure in a slice upstream and a slice downstream the \acrshort{av}:
\begin{equation*}
    \mathrm{TPG}(t) = \frac{1}{|\gammadown|} \int_{\gammadown} \pf(\x, t) \dx - \frac{1}{|\gammaup|} \int_{\gammaup} \pf(\x, t) \dx.
\end{equation*}
We report the \acrshort{tke} and the \acrshort{tpg} for the five patients in \Cref{fig:rt-tpg-tke}. By averaging these results among the five patients, we get  an average  \acrshort{tke} peak of \SI{15.8}{\joule\per\meter\cubed} and an average  \acrshort{tpg} peak of around \SI{13}{mmHg}.

Finally, in the context of \acrshort{as} and \acrshort{tavr}, assessing the probability of thrombus formation is crucial. Regions with elevated thrombotic risk can be identified by analyzing the residence time of the blood: longer residence times correlate with heightened thrombotic risk \cite{rossini2016clinical}. To compute the residence time, we employ an Eulerian approach by advecting  a passive scalar $\Tres(\x, t)$ in the fluid \cite{oks2022fluid}:
\begin{align*}
    & \pdv{\Tres}{t} + \div \left ( \uf \Tres \right ) = 1 & \text{in }  \omegaf \times (0, T), \\
    & \Tres = 0 &  \text{on }  \gammafin \times (0, T), \\
    & \grad \Tres \cdot \nf = 0 &  \text{on }  \gammafw \cup \gammafout \times (0, T).
\end{align*}

\begin{figure}
    \centering
    \includegraphics[trim = {0cm 9.5cm 0cm 0cm}, clip, width=\textwidth]{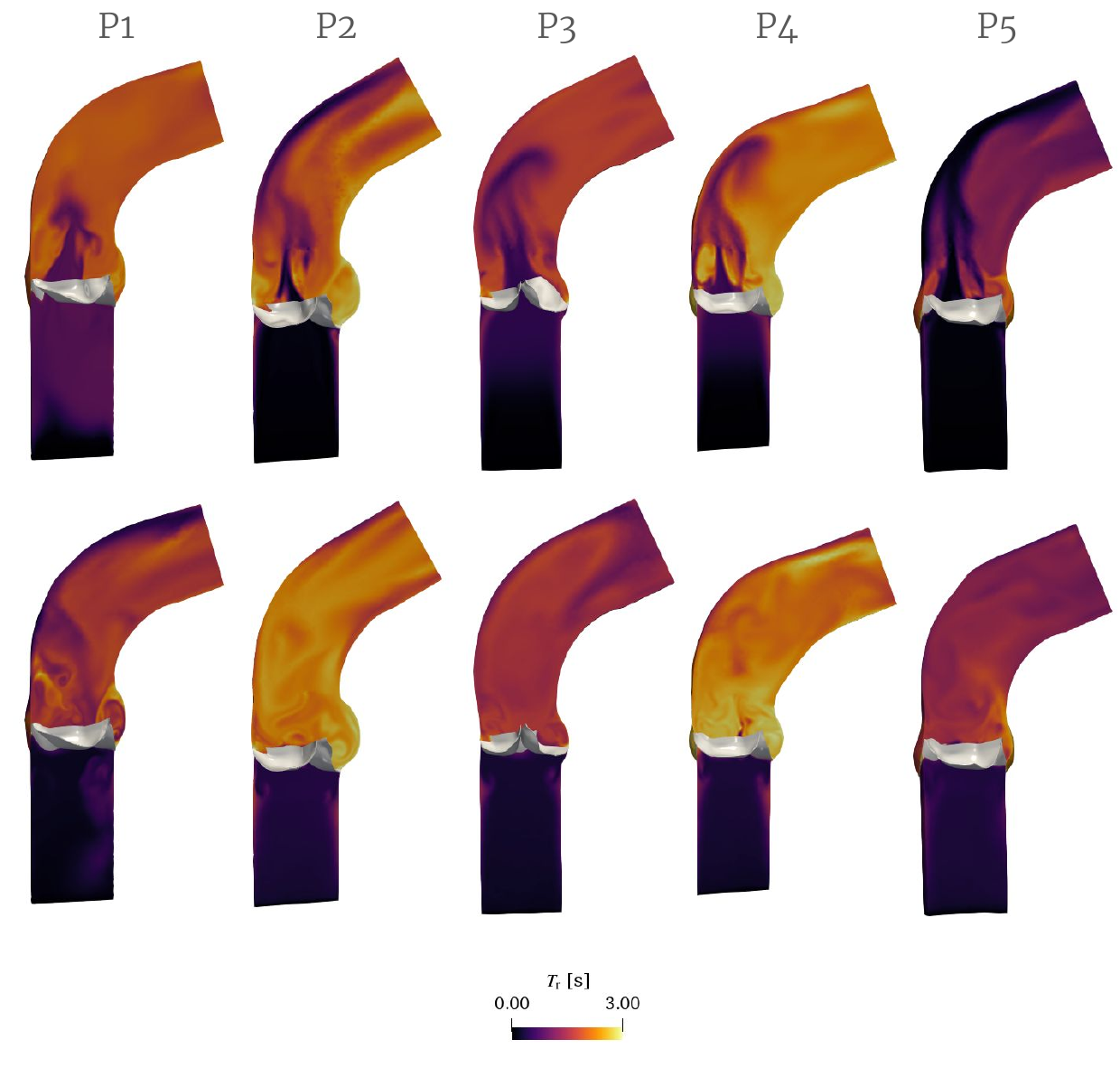}
    \caption{Residence time $\Tres$ for the five patients from \acrshort{fsi} simulations for the five patients: on the top, systole; on the bottom, diastole. Result computed on the last heartbeat.}
    \label{fig:residence-time}
\end{figure}

\Cref{fig:residence-time} illustrates the residence time ($\Tres$) for the five patients during systole and diastole. In both phases of the cardiac cycle, we observe an increase in residence time in the region of \acrshort{sov}, potentially leading to reduced blood washout and increased thrombogenic potential. Consequently, variations in the washout of sinus flow and the presence of stasis regions among different patient models -- described by the residence time -- may contribute to varying risks of leaflet thrombosis and/or valve's leaflets degeneration \cite{hatoum2019impact}. In \Cref{fig:rt-tpg-tke}, we report the residence time averaged in the \acrshort{roi}: after a few cycles, the solution becomes periodic and particles stay in the \acrshort{roi} for about \SI{1.8}{\second} (by averaging the solution among all the patients).

\section{Discussion}
\label{sec:discussion}

We presented a study aimed at validating the blood downstream of the \acrshort{av} using \acrshort{fsi} computational modeling against \fourd{} data. 

Upon examining the \fourd{} data, we observed significant variations in flow rate between different sections. While it was expected for the aorta to exhibit varying flow rates due to its compliance, the magnitude of these variations seemed excessive, considering the standard compliance of the aorta, which typically results in approximately a 10\% enlargement of its diameter \cite{huang2019comparison}. Additionally, upon computing the divergence of the 4D velocity, we observed a significant deviation from the incompressibility constraint. The accuracy of \fourd{} acquisition strongly depends on several parameters such as spatial resolution, signal-to-noise ratio, and non-Gaussian intravoxel velocity distributions \cite{binter2016accuracy, knobloch2014mapping}. In addition, signal loss near the \acrshort{av} compromised the accuracy of these data in this region, making the validation process more challenging \cite{binter2016accuracy}. 

These discrepancies suggested the presence of substantial noise in the \fourd{} acquisition process. To incorporate this information into the computational model, we utilized the average blood flow rate. However, it was crucial to account for this statistical uncertainty in the results. As a matter of fact, validation against \invivo{} data could not be treated as validation against a single "true value"; instead, it was essential for us to establish ranges in order to be able to compare against the distribution of measured kinetic energies. We addressed this by propagating the uncertainty into the kinetic energy within the \acrshort{roi}.

Another critical aspect pertains to the selection of model parameters. We determined that accurately choosing the Young's modulus of the valve is paramount, as this parameter is patient-specific and significantly influences the results. Given its unknown value a priori, we conducted a calibration study to establish a range of variation for valve stiffness that yields outcomes in line with the \invivo{} data and their variation. 

Given the inability to directly measure the valve's Young modulus, lacking precise knowledge of calcification levels, and uniform severity of \acrshort{as} across all patients, we standardized the Young modulus value for the entire cohort. The chosen $\young$ value, calibrated from P1, facilitated comparative analysis across patients without overfitting the model. Moreover, this decision streamlined the pipeline, reducing computational overhead.

Validation of \acrshort{mke} reveals high accuracy for patient P1, with a relative error of 2.5\% in terms of peak kinetic energy. However, for the remaining patients, all \acrshort{mke} values fall within the range of uncertainties, except for patient P2 and P5, where the \acrshort{fsi} simulation underestimates the \invivo{} data. This underscores the significance of careful model parameter selection. Despite the calibration process being conducted on a single patient to establish a range of valve stiffness variation, the majority of other patients' results still fall within our range of uncertainties.

Regarding the validation of blood flow patterns, we observed that, overall, the \acrshort{fsi} model realistically reproduces the patterns observed in the \fourd{} data. This includes the formation of a strong jet through the \acrshort{av} leaflets that impinges on the aortic wall. However, in the case of patient P3, we noted a distinct pattern compared to the other patients: the jet impinged on the opposite side of the wall. Our computational model failed to capture this variation. We attribute this discrepancy to the usage of the valve geometry from P1, adapted for the entire cohort, and potential differences in valve calcification among patients. Calcification could significantly alter the dynamics of \acrshort{av} opening, such as by blocking a specific leaflet. This underscores the importance of incorporating additional patient-specific features, such as valve geometry and calcification, into the computational model.

A comparison of the \acrshort{ava} highlighted how the \acrshort{fsi} results were much more aligned with the measurements obtained from \acrshort{ct} imaging compared to standard \acrshort{ava} measurements through echocardiography. This emphasized how standard methods of \acrshort{ava} measurement, based on the continuity equation, could produce results inconsistent with \acrshort{ct} acquisitions and \insilico{} technologies. This discrepancy mainly stemmed from the numerous assumptions made to compute \acrshort{ava} using the continuity method and the noise and uncertainty associated with the measured quantities used in its computation, which could significantly impact the final \acrshort{ava} result.

In attempting to validate shear stress quantities, we encountered greater challenges. Both computational and \invivo{} results exhibited similar profiles of both \acrshort{sss} and \acrshort{wss}, in terms of spatial distribution and transients. However, discrepancies in magnitudes often arose. The estimation of shear stresses is known to be limited by the relatively low spatial resolution of \fourd{} \cite{petersson2012assessment}. Additionally, we observed very large and unrealistic velocity values at the aorta boundary in the 4D flow data, on the order of \SI{1500}{\milli\meter\per\second}. Consequently, while similar patterns may emerge, the reliability of 4D flow MRI data to this extent is questionable. Moreover, we speculate that incorporating the compliance of the wall into our model (using an \acrshort{fsi} model for blood-wall interaction) would likely result in even smaller \acrshort{wss} values, as supported by recent studies \cite{liu2020fluid}. Therefore, even by enriching our computational model with additional features such as aortic compliance, the \acrshort{wss} results would likely still not align well with those from 4D flow acquisition.

In a similar way, our comparison of turbulence quantities revealed that the computational model qualitatively reproduces the vorticity distribution in the aorta. However, when integrating this quantity to derive blood enstrophy, the \acrshort{fsi} consistently underestimates the \fourd{}. We argue this discrepancy might be attributed to the reliability of \fourd{} data to accurately estimate velocity gradients. 

Finally, we showed how the \acrshort{fsi} simulations allow to compute additional \acrshort{qois} often difficult to acquire with \invivo{} measurements, therefore enlarging the set of metrics and information extracted from medical imaging. This includes residence time, \acrshort{tpg} and \acrshort{tke}. 

\section{Limitations and further developments}
\label{sec:limitations}

% aortic compliance and pressure BC
From a computational modeling perspective, it would be insightful to explore the impact of aortic compliance by incorporating the interaction between blood and the vessel wall, potentially through an Arbitrary Lagrangian Eulerian approach. It can be expected for compliance to affect blood velocity and \acrshort{wss}, potentially resulting in lower values of the latter \cite{liu2020fluid}. In addition, a potential improvement would consist on using a more realistic pressure boundary condition on the outlet section, for instance with an \invivo{} pressure wave or a Windkessel model.

% material model of AV
We utilized a Neo-Hookean constitutive law to model the \acrshort{av}, while anisotropic laws with a fiber description may be more appropriate \cite{cai2021comparison}. However, it is important to note that this choice becomes particularly relevant if validation of valve motion itself is feasible. In this study, our focus was primarily on validating blood dynamics downstream of the valve. In addition, a possible improvement is to model the \acrshort{av} including calcification, for instance using a material model with a higher rigidity. 

% calibration on P1 only
We used the same valve geometry for all the patients, whereas using the patient-specific valve -- ideally directly from \invivo{} measurements -- would potentially affect the results, especially in terms of flow patterns. However, while using the same valve model undoubtedly involves a compromise in terms of accuracy, it strategically balances precision against the preprocessing costs, aligning with the broader objective of our study, and is also in line with other computational studies on this topic \cite{fumagalli2023fluid, criseo2024computational}. Futhermore, our calibration study of the Young modulus was limited to the first patient, and we applied the same valve model to the entire cohort. This approach may introduce biases and influence the outcomes of our results. 
Further investigations with a broader calibration study and patient-specific valve modeling could provide more robust insights.

%4Dflow
Another limitation consists of the coarse resolution of the \fourd{} data: a finer resolution would allow us to carry on a more fair comparison between \insilico{} results and \invivo{} measurements, especially in terms of vorticity. Additionally, signal loss near the valve during the acquisition process compromised the accuracy of the \fourd{} velocity acquisition in that region \cite{binter2016accuracy}.

\section{Conclusions}
\label{sec:conclusions}

In this paper, we presented a comprehensive study aimed at validating blood flow downstream of the \acrshort{av} for five patients with \acrshort{as} using \acrshort{fsi} models, with validation conducted against \fourd{} data.

We introduced a complete pipeline that incorporates various types of \invivo{} data (\fourd{} and \acrshort{ct}), \invitro{} data, and \acrshort{fsi} computational modeling. This approach enabled the development of patient-specific \acrshort{fsi} models for the interaction between the \acrshort{av} and aorta. Our computational model utilizes state-of-the-art \acrshort{ib}-\acrshort{fsi} methods within a \acrshort{hpc} framework.

By utilizing flow rate data computed from 4D flow, we established ranges of variation for kinetic energy within an \acrshort{roi} downstream of the \acrshort{av}. The computational model consistently demonstrated the ability to accurately describe flow patterns and kinetic energy for most patients. Furthermore, our computational results yielded \acrshort{ava} values consistent with those computed using \fourd{}. Notably, both sets of results significantly differed from echocardiography measurements of \acrshort{ava} via the continuity equation.

Additionally, the computation of supplementary quantities derived from phase-averaged velocities, such as shear stresses and vorticity, revealed that while the model accurately captured similar patterns and transients as observed \invivo{}, there were often discrepancies in the values obtained. These discrepancies were attributed to a combination of low resolution and large, unrealistic velocities at the boundary of the aorta characterizing the \fourd{} data.

To summarize, this investigation underscores the efficacy of \acrshort{fsi} models in faithfully replicating \invivo{} dynamics, while identifying the accurate reproduction of crucial indicators within specific ranges of uncertainties. Furthermore, we demonstrated how \acrshort{fsi} enables the computation of supplementary biomechanical parameters, thereby enriching the insights obtainable from medical imaging. In addition, this study illustrates how \insilico{} models can serve as a valuable cross-check to reduce noise and erratic behaviour in \invivo{} data.

\section*{Acknowledgements}
This project was funded by the European Union - EIC Project No 190134524: ``ELEM Virtual Heart Populations for Supercomputers'' (ELVIS). Views and opinions expressed are, however, those of the authors only and do not necessarily reflect those of the European Union or EISMEA. Neither the European Union nor the granting authority can be held responsible for them.

\bibliographystyle{elsarticle-num-names} 

\bibliography{cas-refs}

\printglossary[type=\acronymtype]

\end{document}

%% file: mycommands.tex
\usepackage{lipsum} 
\usepackage[acronym]{glossaries}
\usepackage{todonotes}
\usepackage{cleveref}
\usepackage{stackengine}[2021-07-15]
\usepackage{physics}
\usepackage{bm}
\usepackage{amsmath,amsfonts}
\usepackage{cases}
\usepackage{siunitx}
\usepackage{subcaption}
\usepackage{placeins}

\newcommand{\insilico}{\textit{in silico}}
\newcommand{\invitro}{\textit{in vitro}}
\newcommand{\invivo}{\textit{in vivo}}
\newcommand{\alya}{\textsc{Alya}}

\newacronym{as}{AS}{Aortic Stenosis}
\newacronym{tavr}{TAVR}{Transcatheter Aortic Valve Replacement}
\newacronym{savr}{SAVR}{Surgical Aortic Valve Replacement}
\newacronym{fsi}{FSI}{Fluid Structure Interaction}
\newacronym{av}{AV}{Aortic Valve}
\newacronym{qois}{QoIs}{Quantities of Interest}
\newacronym{ava}{AVA}{Aortic Valve Area}
\newacronym{tpg}{TPG}{Transvalvular Pressure Gradient}
\newacronym{lvot}{LVOT}{Left Ventricle Outflow Track}
\newacronym{cmr}{CMR}{Cardiac Magnetic Resonance}
\newacronym{ct}{CT}{Computed Tomography}
\newacronym{mri}{MRI}{Magnetic Resonance Imaging}
\newacronym{cfd}{CFD}{Computational Fluid Dynamics}
\newacronym{piv}{PIV}{Particle Image Velocimetry}
\newacronym{vvuq}{VVUQ}{Verification, Validation, and Uncertainty Quantification}
\newacronym{fda}{FDA}{Food and Drug Administration}
\newacronym{ib}{IB}{Immersed Boundary}
\newacronym{fe}{FE}{Finite Element}
\newacronym{les}{LES}{Large Eddy Simulation}
\newacronym{ueabs}{UEABS}{Unified European Applications Benchmark Suite}
\newacronym{hpc}{HPC}{High Performance Computing}
\newacronym{roi}{ROI}{Region of Interest}
\newacronym{sov}{SoV}{Sinus of Valsalva}
\newacronym{dofs}{DOFs}{Degrees of Freedom}
\newacronym{mke}{MKE}{Mean Kinetic Energy}
\newacronym{wss}{WSS}{Wall Shear Stress}
\newacronym{tawss}{TAWSS}{Time-Averaged Wall Shear Stress}
\newacronym{sss}{SSS}{Scalar Shear Stress}
\newacronym{lic}{LIC}{Line Integral Convolution}
\newacronym{tke}{TKE}{Turbulent Kinetic Energy}
\newacronym{goa}{GOA}{Geometric Orifice Area}

\makeglossaries

\newcommand{\omegaf}{\Omega_\mathrm{f}}
\newcommand{\omegas}{\Omega_\mathrm{s}}
\newcommand{\gammafin}{\Gamma_\mathrm{f}^\mathrm{in}}
\newcommand{\gammafout}{\Gamma_\mathrm{f}^\mathrm{out}}
\newcommand{\gammafw}{\Gamma_\mathrm{f}^\mathrm{w}}

\newcommand{\gammasw}{\Gamma_\mathrm{s}^\mathrm{w}}
\newcommand{\gammaswet}{\Gamma_\mathrm{s}^\mathrm{wet}}

\newcommand{\rhof}{\rho_\mathrm{f}}
\newcommand{\muf}{\mu_\mathrm{f}}
\newcommand{\uf}{\bm u_\mathrm{f}}
\newcommand{\pf}{p_\mathrm{f}}

\newcommand{\stressf}{{\sigma}_\mathrm{f}}
\newcommand{\tauf}{{\tau_\mathrm{f}}}
\newcommand{\forcing}{\bm f}
\newcommand{\ufin}{\uf^\mathrm{in}}
\newcommand{\nf}{\bm n_\mathrm{f}}
\newcommand{\x}{\bm x}
\newcommand{\Rin}{R_\mathrm{in}}
\newcommand{\uphase}{\langle \uf \rangle}

\newcommand{\displ}{\bm d_\mathrm{s}}
\newcommand{\rhos}{\rho_\mathrm{s}}
\newcommand{\rhoszero}{\rho_\mathrm{s}^0}
\newcommand{\X}{\bm X}
\newcommand{\omegaszero}{\omegas^0}
\newcommand{\divzero}{\grad_0 \cdot}
\newcommand{\young}{E_\mathrm{s}}
\newcommand{\poisson}{\nu_\mathrm{s}}

\newcommand{\dirac}{\bm{\delta}}

\newcommand{\h}[1]{$h_\mathrm{#1}$}

\newcommand{\Thb}{T_\mathrm{HB}}
\newcommand{\Omegaroi}{\omegaf^\mathrm{RoI}}

\newcommand{\mke}{\mathrm{MKE}}
\newcommand{\dx}{\, \mathrm{d}\bm x}
\newcommand{\mkeup}{\mathrm{MKE}^{\mathrm{up}}}
\newcommand{\mkedown}{\mathrm{MKE}^{\mathrm{down}}}
\newcommand{\deltaup}{\delta^{\mathrm{up}}}
\newcommand{\deltadown}{\delta^{\mathrm{down}}}
\newcommand{\mkeavgtilde}{\widetilde{\mathrm{MKE}}^{\mathrm{avg}}}
\newcommand{\mkemintilde}{\widetilde{\mathrm{MKE}}^{\mathrm{min}}}
\newcommand{\mkemaxtilde}{\widetilde{\mathrm{MKE}}^{\mathrm{max}}}
\newcommand{\wss}{\textbf{WSS}}

\newcommand{\sss}{\text{SSS}}
\newcommand{\vorticity}{\bm {\omega}_\mathrm{f}}
\newcommand{\gammaup}{\Gamma_\mathrm{up}}
\newcommand{\gammadown}{\Gamma_\mathrm{down}}
\newcommand{\Tres}{T_\mathrm{r}}

\newcommand{\fourd}{4D flow \acrshort{mri}}
\newcommand{\Qfourdgamma}{Q_{{\Gamma}_j}^\mathrm{4Dflow}}
\newcommand{\Qin}{Q_\mathrm{in}}
\newcommand{\Qfourdavg}{Q_{\mathrm{avg}}^\mathrm{4Dflow}}
\newcommand{\Qfourdmin}{Q_{\mathrm{min}}^\mathrm{4Dflow}}
\newcommand{\Qfourdmax}{Q_{\mathrm{max}}^\mathrm{4Dflow}}
\newcommand{\velfourd}{\bm u^\mathrm{4Dflow}}

\newcommand{\uavg}{\tilde{\bm u}_\mathrm{avg}}
\newcommand{\umax}{\tilde{\bm u}_\mathrm{max}}
\newcommand{\umin}{\tilde{\bm u}_\mathrm{min}}

\newcommand{\Amin}{A_{Q_\mathrm{min}}}
\newcommand{\Amax}{A_{Q_\mathrm{max}}}
\newcommand{\Aavg}{A_\mathrm{avg}}

\DeclareSIUnit\fifthed{^5}
\DeclareSIUnit\millimetermercury{mmHg}

%% file: zingaro-tavr-arxiv.bbl
\begin{thebibliography}{81}
\expandafter\ifx\csname natexlab\endcsname\relax\def\natexlab#1{#1}\fi
\providecommand{\url}[1]{\texttt{#1}}
\providecommand{\href}[2]{#2}
\providecommand{\path}[1]{#1}
\providecommand{\DOIprefix}{doi:}
\providecommand{\ArXivprefix}{arXiv:}
\providecommand{\URLprefix}{URL: }
\providecommand{\Pubmedprefix}{pmid:}
\providecommand{\doi}[1]{\href{http://dx.doi.org/#1}{\path{#1}}}
\providecommand{\Pubmed}[1]{\href{pmid:#1}{\path{#1}}}
\providecommand{\bibinfo}[2]{#2}
\ifx\xfnm\relax \def\xfnm[#1]{\unskip,\space#1}\fi
%Type = Article
\bibitem[{Aluru et~al.(2022)Aluru, Barsouk, Saginala, Rawla, and Barsouk}]{aluru2022valvular}
\bibinfo{author}{J.~S. Aluru}, \bibinfo{author}{A.~Barsouk}, \bibinfo{author}{K.~Saginala}, \bibinfo{author}{P.~Rawla}, \bibinfo{author}{A.~Barsouk},
\newblock \bibinfo{title}{Valvular heart disease epidemiology},
\newblock \bibinfo{journal}{Medical Sciences} \bibinfo{volume}{10} (\bibinfo{year}{2022}) \bibinfo{pages}{32}.
%Type = Misc
\bibitem[{{American Heart Association}(1 08)}]{aha_as}
\bibinfo{author}{{American Heart Association}}, \bibinfo{title}{Aortic stenosis overview}, \bibinfo{howpublished}{\url{https://www.heart.org/en/health-topics/heart-valve-problems-and-disease/heart-valve-problems-and-causes/problem-aortic-valve-stenosis}}, \bibinfo{year}{2024-01-08}.
%Type = Misc
\bibitem[{{National Heart, Lung and Blood Institute}(1 22)}]{nih_tavr}
\bibinfo{author}{{National Heart, Lung and Blood Institute}}, \bibinfo{title}{Transcatheter aortic valve replacement (tavr)}, \bibinfo{howpublished}{\url{https://www.nhlbi.nih.gov/health/tavr}}, \bibinfo{year}{2024-01-22}.
%Type = Article
\bibitem[{Cribier et~al.(2002)Cribier, Eltchaninoff, Bash, Borenstein, Tron, Bauer, Derumeaux, Anselme, Laborde, and Leon}]{cribier2002percutaneous}
\bibinfo{author}{A.~Cribier}, \bibinfo{author}{H.~Eltchaninoff}, \bibinfo{author}{A.~Bash}, \bibinfo{author}{N.~Borenstein}, \bibinfo{author}{C.~Tron}, \bibinfo{author}{F.~Bauer}, \bibinfo{author}{G.~Derumeaux}, \bibinfo{author}{F.~Anselme}, \bibinfo{author}{F.~Laborde}, \bibinfo{author}{M.~B. Leon},
\newblock \bibinfo{title}{Percutaneous transcatheter implantation of an aortic valve prosthesis for calcific aortic stenosis: first human case description},
\newblock \bibinfo{journal}{Circulation} \bibinfo{volume}{106} (\bibinfo{year}{2002}) \bibinfo{pages}{3006--3008}.
%Type = Article
\bibitem[{Fanning et~al.(2013)Fanning, Platts, Walters, and Fraser}]{fanning2013transcatheter}
\bibinfo{author}{J.~P. Fanning}, \bibinfo{author}{D.~G. Platts}, \bibinfo{author}{D.~L. Walters}, \bibinfo{author}{J.~F. Fraser},
\newblock \bibinfo{title}{Transcatheter aortic valve implantation (tavi): valve design and evolution},
\newblock \bibinfo{journal}{International journal of cardiology} \bibinfo{volume}{168} (\bibinfo{year}{2013}) \bibinfo{pages}{1822--1831}.
%Type = Article
\bibitem[{Baumgartner et~al.(2009)Baumgartner, Hung, Bermejo, Chambers, Evangelista, Griffin, Iung, Otto, Pellikka, and Qui{\~n}ones}]{baumgartner2009echocardiographic}
\bibinfo{author}{H.~Baumgartner}, \bibinfo{author}{J.~Hung}, \bibinfo{author}{J.~Bermejo}, \bibinfo{author}{J.~B. Chambers}, \bibinfo{author}{A.~Evangelista}, \bibinfo{author}{B.~P. Griffin}, \bibinfo{author}{B.~Iung}, \bibinfo{author}{C.~M. Otto}, \bibinfo{author}{P.~A. Pellikka}, \bibinfo{author}{M.~Qui{\~n}ones},
\newblock \bibinfo{title}{Echocardiographic assessment of valve stenosis: {EAE/ASE} recommendations for clinical practice},
\newblock \bibinfo{journal}{European Journal of Echocardiography} \bibinfo{volume}{10} (\bibinfo{year}{2009}) \bibinfo{pages}{1--25}.
%Type = Article
\bibitem[{Otto et~al.(1986)Otto, Pearlman, Comess, Reamer, Janko, and Huntsman}]{otto1986determination}
\bibinfo{author}{C.~M. Otto}, \bibinfo{author}{A.~S. Pearlman}, \bibinfo{author}{K.~A. Comess}, \bibinfo{author}{R.~P. Reamer}, \bibinfo{author}{C.~L. Janko}, \bibinfo{author}{L.~L. Huntsman},
\newblock \bibinfo{title}{Determination of the stenotic aortic valve area in adults using doppler echocardiography},
\newblock \bibinfo{journal}{journal of the American College of Cardiology} \bibinfo{volume}{7} (\bibinfo{year}{1986}) \bibinfo{pages}{509--517}.
%Type = Article
\bibitem[{Zoghbi et~al.(2009)Zoghbi, Chambers, Dumesnil, Foster, Gottdiener, Grayburn, Khandheria, Levine, Marx, Miller et~al.}]{zoghbi2009recommendations}
\bibinfo{author}{W.~A. Zoghbi}, \bibinfo{author}{J.~B. Chambers}, \bibinfo{author}{J.~G. Dumesnil}, \bibinfo{author}{E.~Foster}, \bibinfo{author}{J.~S. Gottdiener}, \bibinfo{author}{P.~A. Grayburn}, \bibinfo{author}{B.~K. Khandheria}, \bibinfo{author}{R.~A. Levine}, \bibinfo{author}{G.~R. Marx}, \bibinfo{author}{F.~A. Miller}, et~al.,
\newblock \bibinfo{title}{Recommendations for evaluation of prosthetic valves with echocardiography and doppler ultrasound: a report from the american society of echocardiography's guidelines and standards committee and the task force on prosthetic valves, developed in conjunction with the american college of cardiology cardiovascular imaging committee, cardiac imaging committee of the american heart association, the european association of echocardiography, a registered branch of the european society of cardiology, the japanese society of echocardiography and the canadian society of echocardiography, endorsed by the american college of cardiology foundation, american heart association, european association of echocardiography, a registered branch of the european society of cardiology, the japanese society of echocardiography, and canadian society of echocardiography},
\newblock \bibinfo{journal}{Journal of the American Society of Echocardiography} \bibinfo{volume}{22} (\bibinfo{year}{2009}) \bibinfo{pages}{975--1014}.
%Type = Article
\bibitem[{Archer et~al.(2020)Archer, Elhawaz, Barker, Fidock, Rothman, Van Der~Geest, Hose, Briffa, Hall, Grech et~al.}]{archer2020validation}
\bibinfo{author}{G.~T. Archer}, \bibinfo{author}{A.~Elhawaz}, \bibinfo{author}{N.~Barker}, \bibinfo{author}{B.~Fidock}, \bibinfo{author}{A.~Rothman}, \bibinfo{author}{R.~Van Der~Geest}, \bibinfo{author}{R.~Hose}, \bibinfo{author}{N.~Briffa}, \bibinfo{author}{I.~R. Hall}, \bibinfo{author}{E.~Grech}, et~al.,
\newblock \bibinfo{title}{Validation of four-dimensional flow cardiovascular magnetic resonance for aortic stenosis assessment},
\newblock \bibinfo{journal}{Scientific reports} \bibinfo{volume}{10} (\bibinfo{year}{2020}) \bibinfo{pages}{10569}.
%Type = Article
\bibitem[{Manzo et~al.(2023)Manzo, Ilardi, Nappa, Mariani, Angellotti, Immobile~Molaro, Sgherzi, Castiello, Simonetti, Santoro et~al.}]{manzo2023echocardiographic}
\bibinfo{author}{R.~Manzo}, \bibinfo{author}{F.~Ilardi}, \bibinfo{author}{D.~Nappa}, \bibinfo{author}{A.~Mariani}, \bibinfo{author}{D.~Angellotti}, \bibinfo{author}{M.~Immobile~Molaro}, \bibinfo{author}{G.~Sgherzi}, \bibinfo{author}{D.~S. Castiello}, \bibinfo{author}{F.~Simonetti}, \bibinfo{author}{C.~Santoro}, et~al.,
\newblock \bibinfo{title}{Echocardiographic evaluation of aortic stenosis: A comprehensive review},
\newblock \bibinfo{journal}{Diagnostics} \bibinfo{volume}{13} (\bibinfo{year}{2023}) \bibinfo{pages}{2527}.
%Type = Article
\bibitem[{Ngo et~al.(2019)Ngo, Kim, Jung, Chung, Lee, and Kwak}]{ngo2019four}
\bibinfo{author}{M.~T. Ngo}, \bibinfo{author}{C.~I. Kim}, \bibinfo{author}{J.~Jung}, \bibinfo{author}{G.~H. Chung}, \bibinfo{author}{D.~H. Lee}, \bibinfo{author}{H.~S. Kwak},
\newblock \bibinfo{title}{Four-dimensional flow magnetic resonance imaging for assessment of velocity magnitudes and flow patterns in the human carotid artery bifurcation: Comparison with computational fluid dynamics},
\newblock \bibinfo{journal}{Diagnostics} \bibinfo{volume}{9} (\bibinfo{year}{2019}) \bibinfo{pages}{223}.
%Type = Article
\bibitem[{Clavel et~al.(2015)Clavel, Malouf, Messika-Zeitoun, Araoz, Michelena, and Enriquez-Sarano}]{clavel2015aortic}
\bibinfo{author}{M.-A. Clavel}, \bibinfo{author}{J.~Malouf}, \bibinfo{author}{D.~Messika-Zeitoun}, \bibinfo{author}{P.~A. Araoz}, \bibinfo{author}{H.~I. Michelena}, \bibinfo{author}{M.~Enriquez-Sarano},
\newblock \bibinfo{title}{Aortic valve area calculation in aortic stenosis by ct and doppler echocardiography},
\newblock \bibinfo{journal}{Cardiovascular Imaging} \bibinfo{volume}{8} (\bibinfo{year}{2015}) \bibinfo{pages}{248--257}.
%Type = Article
\bibitem[{Velders et~al.(2022)Velders, Groenwold, Kappetein, Braun, Klautz, and Vriesendorp}]{velders2022measurement}
\bibinfo{author}{B.~Velders}, \bibinfo{author}{R.~Groenwold}, \bibinfo{author}{A.~Kappetein}, \bibinfo{author}{J.~Braun}, \bibinfo{author}{R.~Klautz}, \bibinfo{author}{M.~Vriesendorp},
\newblock \bibinfo{title}{Measurement error in echocardiographic assessment of aortic stenosis: an epidemiological consideration of research methodology and clinical practice},
\newblock \bibinfo{journal}{European Heart Journal} \bibinfo{volume}{43} (\bibinfo{year}{2022}) \bibinfo{pages}{ehac544--2863}.
%Type = Article
\bibitem[{Bahraseman et~al.(2016)Bahraseman, Languri, Yahyapourjalaly, and Espino}]{bahraseman2016fluid}
\bibinfo{author}{H.~G. Bahraseman}, \bibinfo{author}{E.~M. Languri}, \bibinfo{author}{N.~Yahyapourjalaly}, \bibinfo{author}{D.~M. Espino},
\newblock \bibinfo{title}{Fluid-structure interaction modeling of aortic valve stenosis at different heart rates},
\newblock \bibinfo{journal}{Acta of bioengineering and biomechanics} \bibinfo{volume}{18} (\bibinfo{year}{2016}) \bibinfo{pages}{11--20}.
%Type = Article
\bibitem[{Zhu et~al.(2022)Zhu, Seo, and Mittal}]{zhu2022computational}
\bibinfo{author}{C.~Zhu}, \bibinfo{author}{J.-H. Seo}, \bibinfo{author}{R.~Mittal},
\newblock \bibinfo{title}{Computational modeling of aortic stenosis with a reduced degree-of-freedom fluid-structure interaction valve model},
\newblock \bibinfo{journal}{Journal of Biomechanical Engineering} \bibinfo{volume}{144} (\bibinfo{year}{2022}) \bibinfo{pages}{031012}.
%Type = Article
\bibitem[{Fumagalli et~al.(2023)Fumagalli, Polidori, Renzi, Fusini, Quarteroni, Pontone, and Vergara}]{fumagalli2023fluid}
\bibinfo{author}{I.~Fumagalli}, \bibinfo{author}{R.~Polidori}, \bibinfo{author}{F.~Renzi}, \bibinfo{author}{L.~Fusini}, \bibinfo{author}{A.~Quarteroni}, \bibinfo{author}{G.~Pontone}, \bibinfo{author}{C.~Vergara},
\newblock \bibinfo{title}{Fluid-structure interaction analysis of transcatheter aortic valve implantation},
\newblock \bibinfo{journal}{International Journal for Numerical Methods in Biomedical Engineering}  (\bibinfo{year}{2023}) \bibinfo{pages}{e3704}.
%Type = Article
\bibitem[{Luraghi et~al.(2019)Luraghi, Migliavacca, Garc{\'\i}a-Gonz{\'a}lez, Chiastra, Rossi, Cao, Stefanini, and Rodriguez~Matas}]{luraghi2019modeling}
\bibinfo{author}{G.~Luraghi}, \bibinfo{author}{F.~Migliavacca}, \bibinfo{author}{A.~Garc{\'\i}a-Gonz{\'a}lez}, \bibinfo{author}{C.~Chiastra}, \bibinfo{author}{A.~Rossi}, \bibinfo{author}{D.~Cao}, \bibinfo{author}{G.~Stefanini}, \bibinfo{author}{J.~F. Rodriguez~Matas},
\newblock \bibinfo{title}{On the modeling of patient-specific transcatheter aortic valve replacement: a fluid--structure interaction approach},
\newblock \bibinfo{journal}{Cardiovascular engineering and technology} \bibinfo{volume}{10} (\bibinfo{year}{2019}) \bibinfo{pages}{437--455}.
%Type = Article
\bibitem[{Luraghi et~al.(2020)Luraghi, Matas, Beretta, Chiozzi, Iannetti, and Migliavacca}]{luraghi2020impact}
\bibinfo{author}{G.~Luraghi}, \bibinfo{author}{J.~F.~R. Matas}, \bibinfo{author}{M.~Beretta}, \bibinfo{author}{N.~Chiozzi}, \bibinfo{author}{L.~Iannetti}, \bibinfo{author}{F.~Migliavacca},
\newblock \bibinfo{title}{The impact of calcification patterns in transcatheter aortic valve performance: a fluid-structure interaction analysis},
\newblock \bibinfo{journal}{Computer Methods in Biomechanics and Biomedical Engineering} \bibinfo{volume}{24} (\bibinfo{year}{2020}) \bibinfo{pages}{375--383}.
%Type = Article
\bibitem[{Ghosh et~al.(2020)Ghosh, Marom, Bianchi, D’souza, Zietak, and Bluestein}]{ghosh2020numerical}
\bibinfo{author}{R.~P. Ghosh}, \bibinfo{author}{G.~Marom}, \bibinfo{author}{M.~Bianchi}, \bibinfo{author}{K.~D’souza}, \bibinfo{author}{W.~Zietak}, \bibinfo{author}{D.~Bluestein},
\newblock \bibinfo{title}{Numerical evaluation of transcatheter aortic valve performance during heart beating and its post-deployment fluid--structure interaction analysis},
\newblock \bibinfo{journal}{Biomechanics and modeling in mechanobiology} \bibinfo{volume}{19} (\bibinfo{year}{2020}) \bibinfo{pages}{1725--1740}.
%Type = Article
\bibitem[{Basri et~al.(2020)Basri, Zuber, Basri, Zakaria, Aziz, Tamagawa, Ahmad et~al.}]{basri2020fluid}
\bibinfo{author}{A.~A. Basri}, \bibinfo{author}{M.~Zuber}, \bibinfo{author}{E.~I. Basri}, \bibinfo{author}{M.~S. Zakaria}, \bibinfo{author}{A.~F. Aziz}, \bibinfo{author}{M.~Tamagawa}, \bibinfo{author}{K.~A. Ahmad}, et~al.,
\newblock \bibinfo{title}{Fluid structure interaction on paravalvular leakage of transcatheter aortic valve implantation related to aortic stenosis: A patient-specific case},
\newblock \bibinfo{journal}{Computational and mathematical methods in medicine} \bibinfo{volume}{2020} (\bibinfo{year}{2020}).
%Type = Article
\bibitem[{De~Hart et~al.(2003)De~Hart, Baaijens, Peters, and Schreurs}]{de2003computational}
\bibinfo{author}{J.~De~Hart}, \bibinfo{author}{F.~Baaijens}, \bibinfo{author}{G.~Peters}, \bibinfo{author}{P.~Schreurs},
\newblock \bibinfo{title}{A computational fluid-structure interaction analysis of a fiber-reinforced stentless aortic valve},
\newblock \bibinfo{journal}{Journal of biomechanics} \bibinfo{volume}{36} (\bibinfo{year}{2003}) \bibinfo{pages}{699--712}.
%Type = Article
\bibitem[{Oks et~al.(2023{\natexlab{a}})Oks, Reza, V{\'a}zquez, Houzeaux, Kovarovic, Samaniego, and Bluestein}]{oks2023effectsino}
\bibinfo{author}{D.~Oks}, \bibinfo{author}{S.~Reza}, \bibinfo{author}{M.~V{\'a}zquez}, \bibinfo{author}{G.~Houzeaux}, \bibinfo{author}{B.~Kovarovic}, \bibinfo{author}{C.~Samaniego}, \bibinfo{author}{D.~Bluestein},
\newblock \bibinfo{title}{Effect of sinotubular junction size on {TAVR} leaflet thrombosis: A fluid--structure interaction analysis},
\newblock \bibinfo{journal}{Annals of Biomedical Engineering}  (\bibinfo{year}{2023}{\natexlab{a}}) \bibinfo{pages}{1--15}.
%Type = Article
\bibitem[{Oks et~al.(2023{\natexlab{b}})Oks, Houzeaux, V{\'a}zquez, Neidlin, and Samaniego}]{oks2023effect}
\bibinfo{author}{D.~Oks}, \bibinfo{author}{G.~Houzeaux}, \bibinfo{author}{M.~V{\'a}zquez}, \bibinfo{author}{M.~Neidlin}, \bibinfo{author}{C.~Samaniego},
\newblock \bibinfo{title}{Effect of {TAVR} commissural alignment on coronary flow: A fluid-structure interaction analysis},
\newblock \bibinfo{journal}{Computer Methods and Programs in Biomedicine} \bibinfo{volume}{242} (\bibinfo{year}{2023}{\natexlab{b}}) \bibinfo{pages}{107818}.
%Type = Article
\bibitem[{Oks et~al.(2022)Oks, Samaniego, Houzeaux, Butakoff, and V{\'a}zquez}]{oks2022fluid}
\bibinfo{author}{D.~Oks}, \bibinfo{author}{C.~Samaniego}, \bibinfo{author}{G.~Houzeaux}, \bibinfo{author}{C.~Butakoff}, \bibinfo{author}{M.~V{\'a}zquez},
\newblock \bibinfo{title}{Fluid--structure interaction analysis of eccentricity and leaflet rigidity on thrombosis biomarkers in bioprosthetic aortic valve replacements},
\newblock \bibinfo{journal}{International Journal for Numerical Methods in Biomedical Engineering} \bibinfo{volume}{38} (\bibinfo{year}{2022}) \bibinfo{pages}{e3649}.
%Type = Article
\bibitem[{McQueen et~al.(1982)McQueen, Peskin, and Yellin}]{mcqueen1982fluid}
\bibinfo{author}{D.~M. McQueen}, \bibinfo{author}{C.~S. Peskin}, \bibinfo{author}{E.~L. Yellin},
\newblock \bibinfo{title}{Fluid dynamics of the mitral valve: physiological aspects of a mathematical model},
\newblock \bibinfo{journal}{American Journal of Physiology-Heart and Circulatory Physiology} \bibinfo{volume}{242} (\bibinfo{year}{1982}) \bibinfo{pages}{H1095--H1110}.
%Type = Article
\bibitem[{Astorino et~al.(2009)Astorino, Gerbeau, Pantz, and Traor{\'e}}]{astorino2009fluid}
\bibinfo{author}{M.~Astorino}, \bibinfo{author}{J.-F. Gerbeau}, \bibinfo{author}{O.~Pantz}, \bibinfo{author}{K.-F. Traor{\'e}},
\newblock \bibinfo{title}{Fluid--structure interaction and multi-body contact: application to aortic valves},
\newblock \bibinfo{journal}{Computer Methods in Applied Mechanics and Engineering} \bibinfo{volume}{198} (\bibinfo{year}{2009}) \bibinfo{pages}{3603--3612}.
%Type = Article
\bibitem[{Griffith(2012)}]{griffith2012immersed}
\bibinfo{author}{B.~E. Griffith},
\newblock \bibinfo{title}{Immersed boundary model of aortic heart valve dynamics with physiological driving and loading conditions},
\newblock \bibinfo{journal}{International journal for numerical methods in biomedical engineering} \bibinfo{volume}{28} (\bibinfo{year}{2012}) \bibinfo{pages}{317--345}.
%Type = Article
\bibitem[{Marom(2015)}]{marom2015numerical}
\bibinfo{author}{G.~Marom},
\newblock \bibinfo{title}{Numerical methods for fluid--structure interaction models of aortic valves},
\newblock \bibinfo{journal}{Archives of Computational Methods in Engineering} \bibinfo{volume}{22} (\bibinfo{year}{2015}) \bibinfo{pages}{595--620}.
%Type = Article
\bibitem[{Aboelkassem et~al.(2015)Aboelkassem, Savic, and Campbell}]{aboelkassem2015mathematical}
\bibinfo{author}{Y.~Aboelkassem}, \bibinfo{author}{D.~Savic}, \bibinfo{author}{S.~G. Campbell},
\newblock \bibinfo{title}{Mathematical modeling of aortic valve dynamics during systole},
\newblock \bibinfo{journal}{Journal of theoretical biology} \bibinfo{volume}{365} (\bibinfo{year}{2015}) \bibinfo{pages}{280--288}.
%Type = Article
\bibitem[{Flamini et~al.(2016)Flamini, DeAnda, and Griffith}]{flamini2016immersed}
\bibinfo{author}{V.~Flamini}, \bibinfo{author}{A.~DeAnda}, \bibinfo{author}{B.~E. Griffith},
\newblock \bibinfo{title}{Immersed boundary-finite element model of fluid--structure interaction in the aortic root},
\newblock \bibinfo{journal}{Theoretical and computational fluid dynamics} \bibinfo{volume}{30} (\bibinfo{year}{2016}) \bibinfo{pages}{139--164}.
%Type = Article
\bibitem[{Fedele et~al.(2017)Fedele, Faggiano, Ded{\`e}, and Quarteroni}]{fedele2017patient}
\bibinfo{author}{M.~Fedele}, \bibinfo{author}{E.~Faggiano}, \bibinfo{author}{L.~Ded{\`e}}, \bibinfo{author}{A.~Quarteroni},
\newblock \bibinfo{title}{A patient-specific aortic valve model based on moving resistive immersed implicit surfaces},
\newblock \bibinfo{journal}{Biomechanics and modeling in mechanobiology} \bibinfo{volume}{16} (\bibinfo{year}{2017}) \bibinfo{pages}{1779--1803}.
%Type = Article
\bibitem[{Sp{\"u}hler et~al.(2018)Sp{\"u}hler, Jansson, Jansson, and Hoffman}]{spuhler20183d}
\bibinfo{author}{J.~H. Sp{\"u}hler}, \bibinfo{author}{J.~Jansson}, \bibinfo{author}{N.~Jansson}, \bibinfo{author}{J.~Hoffman},
\newblock \bibinfo{title}{3d fluid-structure interaction simulation of aortic valves using a unified continuum ale fem model},
\newblock \bibinfo{journal}{Frontiers in physiology} \bibinfo{volume}{9} (\bibinfo{year}{2018}) \bibinfo{pages}{363}.
%Type = Article
\bibitem[{Wu et~al.(2019)Wu, Muchowski, Johnson, Rajanna, and Hsu}]{wu2019immersogeometric}
\bibinfo{author}{M.~C. Wu}, \bibinfo{author}{H.~M. Muchowski}, \bibinfo{author}{E.~L. Johnson}, \bibinfo{author}{M.~R. Rajanna}, \bibinfo{author}{M.-C. Hsu},
\newblock \bibinfo{title}{Immersogeometric fluid--structure interaction modeling and simulation of transcatheter aortic valve replacement},
\newblock \bibinfo{journal}{Computer Methods in Applied Mechanics and Engineering} \bibinfo{volume}{357} (\bibinfo{year}{2019}) \bibinfo{pages}{112556}.
%Type = Article
\bibitem[{Fumagalli et~al.(2020)Fumagalli, Fedele, Vergara, Ippolito, Nicol{\`o}, Antona, Scrofani, Quarteroni et~al.}]{fumagalli2020image}
\bibinfo{author}{I.~Fumagalli}, \bibinfo{author}{M.~Fedele}, \bibinfo{author}{C.~Vergara}, \bibinfo{author}{S.~Ippolito}, \bibinfo{author}{F.~Nicol{\`o}}, \bibinfo{author}{C.~Antona}, \bibinfo{author}{R.~Scrofani}, \bibinfo{author}{A.~Quarteroni}, et~al.,
\newblock \bibinfo{title}{An image-based computational hemodynamics study of the systolic anterior motion of the mitral valve},
\newblock \bibinfo{journal}{Computers in biology and medicine} \bibinfo{volume}{123} (\bibinfo{year}{2020}) \bibinfo{pages}{103922}.
%Type = Article
\bibitem[{Zingaro et~al.(2022)Zingaro, Bucelli, Fumagalli, Dede', and Quarteroni}]{zingaro2022modeling}
\bibinfo{author}{A.~Zingaro}, \bibinfo{author}{M.~Bucelli}, \bibinfo{author}{I.~Fumagalli}, \bibinfo{author}{L.~Dede'}, \bibinfo{author}{A.~Quarteroni},
\newblock \bibinfo{title}{Modeling isovolumetric phases in cardiac flows by an augmented resistive immersed implicit surface method},
\newblock \bibinfo{journal}{International Journal for Numerical Methods in Biomedical Engineering}  (\bibinfo{year}{2022}) \bibinfo{pages}{e3767}.
%Type = Article
\bibitem[{Bennati et~al.(2023)Bennati, Giambruno, Renzi, Di~Nicola, Maffeis, Puppini, Luciani, and Vergara}]{bennati2023turbulent}
\bibinfo{author}{L.~Bennati}, \bibinfo{author}{V.~Giambruno}, \bibinfo{author}{F.~Renzi}, \bibinfo{author}{V.~Di~Nicola}, \bibinfo{author}{C.~Maffeis}, \bibinfo{author}{G.~Puppini}, \bibinfo{author}{G.~B. Luciani}, \bibinfo{author}{C.~Vergara},
\newblock \bibinfo{title}{Turbulent blood dynamics in the left heart in the presence of mitral regurgitation: a computational study based on multi-series cine-mri},
\newblock \bibinfo{journal}{Biomechanics and Modeling in Mechanobiology} \bibinfo{volume}{22} (\bibinfo{year}{2023}) \bibinfo{pages}{1829--1846}.
%Type = Article
\bibitem[{Zingaro et~al.(2024)Zingaro, Bucelli, Piersanti, Regazzoni, Dede', and Quarteroni}]{zingaro2024electromechanics}
\bibinfo{author}{A.~Zingaro}, \bibinfo{author}{M.~Bucelli}, \bibinfo{author}{R.~Piersanti}, \bibinfo{author}{F.~Regazzoni}, \bibinfo{author}{L.~Dede'}, \bibinfo{author}{A.~Quarteroni},
\newblock \bibinfo{title}{An electromechanics-driven fluid dynamics model for the simulation of the whole human heart},
\newblock \bibinfo{journal}{Journal of Computational Physics}  (\bibinfo{year}{2024}) \bibinfo{pages}{112885}.
%Type = Article
\bibitem[{Kuchumov et~al.(2023)Kuchumov, Makashova, Vladimirov, Borodin, and Dokuchaeva}]{kuchumov2023fluid}
\bibinfo{author}{A.~G. Kuchumov}, \bibinfo{author}{A.~Makashova}, \bibinfo{author}{S.~Vladimirov}, \bibinfo{author}{V.~Borodin}, \bibinfo{author}{A.~Dokuchaeva},
\newblock \bibinfo{title}{Fluid--structure interaction aortic valve surgery simulation: A review},
\newblock \bibinfo{journal}{Fluids} \bibinfo{volume}{8} (\bibinfo{year}{2023}) \bibinfo{pages}{295}.
%Type = Techreport
\bibitem[{{U.S. Department of Health and Human Services} et~al.(2023){U.S. Department of Health and Human Services}, {Food and Drug Administration}, and {Center for Devices and Radiological Health}}]{fda_vvuq}
\bibinfo{author}{{U.S. Department of Health and Human Services}}, \bibinfo{author}{{Food and Drug Administration}}, \bibinfo{author}{{Center for Devices and Radiological Health}}, \bibinfo{title}{Assessing the Credibility of Computational Modeling and Simulation in Medical Device Submissions}, \bibinfo{type}{Technical Report}, \bibinfo{year}{2023}.
%Type = Article
\bibitem[{Gallo et~al.(2014)Gallo, G{\"u}lan, Di~Stefano, Ponzini, L{\"u}thi, Holzner, and Morbiducci}]{gallo2014analysis}
\bibinfo{author}{D.~Gallo}, \bibinfo{author}{U.~G{\"u}lan}, \bibinfo{author}{A.~Di~Stefano}, \bibinfo{author}{R.~Ponzini}, \bibinfo{author}{B.~L{\"u}thi}, \bibinfo{author}{M.~Holzner}, \bibinfo{author}{U.~Morbiducci},
\newblock \bibinfo{title}{Analysis of thoracic aorta hemodynamics using 3d particle tracking velocimetry and computational fluid dynamics},
\newblock \bibinfo{journal}{Journal of biomechanics} \bibinfo{volume}{47} (\bibinfo{year}{2014}) \bibinfo{pages}{3149--3155}.
%Type = Article
\bibitem[{Toggweiler et~al.(2022)Toggweiler, De~Boeck, Karakas, and G{\"u}lan}]{toggweiler2022turbulent}
\bibinfo{author}{S.~Toggweiler}, \bibinfo{author}{B.~De~Boeck}, \bibinfo{author}{O.~Karakas}, \bibinfo{author}{U.~G{\"u}lan},
\newblock \bibinfo{title}{Turbulent kinetic energy loss and shear stresses before and after transcatheter aortic valve replacement},
\newblock \bibinfo{journal}{Case Reports} \bibinfo{volume}{4} (\bibinfo{year}{2022}) \bibinfo{pages}{318--320}.
%Type = Article
\bibitem[{G{\"u}lan et~al.(2022)G{\"u}lan, Rossi, Gotschy, Saguner, Manka, Brunckhorst, Duru, Schmied, and Niederseer}]{gulan2022comparative}
\bibinfo{author}{U.~G{\"u}lan}, \bibinfo{author}{V.~A. Rossi}, \bibinfo{author}{A.~Gotschy}, \bibinfo{author}{A.~M. Saguner}, \bibinfo{author}{R.~Manka}, \bibinfo{author}{C.~B. Brunckhorst}, \bibinfo{author}{F.~Duru}, \bibinfo{author}{C.~M. Schmied}, \bibinfo{author}{D.~Niederseer},
\newblock \bibinfo{title}{A comparative study on the analysis of hemodynamics in the athlete’s heart},
\newblock \bibinfo{journal}{Scientific Reports} \bibinfo{volume}{12} (\bibinfo{year}{2022}) \bibinfo{pages}{16666}.
%Type = Article
\bibitem[{Dumont et~al.(2004)Dumont, Stijnen, Vierendeels, Van De~Vosse, and Verdonck}]{dumont2004validation}
\bibinfo{author}{K.~Dumont}, \bibinfo{author}{J.~Stijnen}, \bibinfo{author}{J.~Vierendeels}, \bibinfo{author}{F.~Van De~Vosse}, \bibinfo{author}{P.~Verdonck},
\newblock \bibinfo{title}{Validation of a fluid--structure interaction model of a heart valve using the dynamic mesh method in fluent},
\newblock \bibinfo{journal}{Computer methods in biomechanics and biomedical engineering} \bibinfo{volume}{7} (\bibinfo{year}{2004}) \bibinfo{pages}{139--146}.
%Type = Article
\bibitem[{Luraghi et~al.(2017)Luraghi, Wu, De~Gaetano, Matas, Moggridge, Serrani, Stasiak, Costantino, and Migliavacca}]{luraghi2017evaluation}
\bibinfo{author}{G.~Luraghi}, \bibinfo{author}{W.~Wu}, \bibinfo{author}{F.~De~Gaetano}, \bibinfo{author}{J.~F.~R. Matas}, \bibinfo{author}{G.~D. Moggridge}, \bibinfo{author}{M.~Serrani}, \bibinfo{author}{J.~Stasiak}, \bibinfo{author}{M.~L. Costantino}, \bibinfo{author}{F.~Migliavacca},
\newblock \bibinfo{title}{Evaluation of an aortic valve prosthesis: Fluid-structure interaction or structural simulation?},
\newblock \bibinfo{journal}{Journal of biomechanics} \bibinfo{volume}{58} (\bibinfo{year}{2017}) \bibinfo{pages}{45--51}.
%Type = Article
\bibitem[{Tango et~al.(2018)Tango, Salmonsmith, Ducci, and Burriesci}]{tango2018validation}
\bibinfo{author}{A.~M. Tango}, \bibinfo{author}{J.~Salmonsmith}, \bibinfo{author}{A.~Ducci}, \bibinfo{author}{G.~Burriesci},
\newblock \bibinfo{title}{Validation and extension of a fluid--structure interaction model of the healthy aortic valve},
\newblock \bibinfo{journal}{Cardiovascular Engineering and Technology} \bibinfo{volume}{9} (\bibinfo{year}{2018}) \bibinfo{pages}{739--751}.
%Type = Article
\bibitem[{Sodhani et~al.(2018)Sodhani, Reese, Aksenov, So{\u{g}}anci, Jockenh{\"o}vel, Mela, and Stapleton}]{sodhani2018fluid}
\bibinfo{author}{D.~Sodhani}, \bibinfo{author}{S.~Reese}, \bibinfo{author}{A.~Aksenov}, \bibinfo{author}{S.~So{\u{g}}anci}, \bibinfo{author}{S.~Jockenh{\"o}vel}, \bibinfo{author}{P.~Mela}, \bibinfo{author}{S.~E. Stapleton},
\newblock \bibinfo{title}{Fluid-structure interaction simulation of artificial textile reinforced aortic heart valve: Validation with an in-vitro test},
\newblock \bibinfo{journal}{Journal of biomechanics} \bibinfo{volume}{78} (\bibinfo{year}{2018}) \bibinfo{pages}{52--69}.
%Type = Inproceedings
\bibitem[{Borowski et~al.(2022)Borowski, Ott, Oldenburg, Kaule, {\"O}ner, Schmitz, and Stiehm}]{borowski2022validation}
\bibinfo{author}{F.~Borowski}, \bibinfo{author}{R.~Ott}, \bibinfo{author}{J.~Oldenburg}, \bibinfo{author}{S.~Kaule}, \bibinfo{author}{A.~{\"O}ner}, \bibinfo{author}{K.-P. Schmitz}, \bibinfo{author}{M.~Stiehm},
\newblock \bibinfo{title}{Validation of a fluid structure interaction model for {TAVR} using particle image velocimetry},
\newblock in: \bibinfo{booktitle}{Current Directions in Biomedical Engineering}, volume~\bibinfo{volume}{8}, \bibinfo{organization}{De Gruyter}, \bibinfo{year}{2022}, pp. \bibinfo{pages}{512--515}.
%Type = Article
\bibitem[{Peskin(1972)}]{peskin1972flow}
\bibinfo{author}{C.~S. Peskin},
\newblock \bibinfo{title}{Flow patterns around heart valves: a numerical method},
\newblock \bibinfo{journal}{Journal of computational physics} \bibinfo{volume}{10} (\bibinfo{year}{1972}) \bibinfo{pages}{252--271}.
%Type = Incollection
\bibitem[{Kikinis et~al.(2013)Kikinis, Pieper, and Vosburgh}]{kikinis20133d}
\bibinfo{author}{R.~Kikinis}, \bibinfo{author}{S.~D. Pieper}, \bibinfo{author}{K.~G. Vosburgh},
\newblock \bibinfo{title}{3d slicer: a platform for subject-specific image analysis, visualization, and clinical support},
\newblock in: \bibinfo{booktitle}{Intraoperative imaging and image-guided therapy}, \bibinfo{publisher}{Springer}, \bibinfo{year}{2013}, pp. \bibinfo{pages}{277--289}.
%Type = Misc
\bibitem[{{Beta Simulation Solution}(1 22)}]{ansa}
\bibinfo{author}{{Beta Simulation Solution}}, \bibinfo{title}{Ansa preprocessor. the advanced cae pre-processing software for complete model build up}, \bibinfo{howpublished}{\url{https://www.beta-cae.com/ansa.htm}}, \bibinfo{year}{2024-01-22}.
%Type = Article
\bibitem[{G{\"u}lan et~al.(2017)G{\"u}lan, Binter, Kozerke, and Holzner}]{gulan2017shear}
\bibinfo{author}{U.~G{\"u}lan}, \bibinfo{author}{C.~Binter}, \bibinfo{author}{S.~Kozerke}, \bibinfo{author}{M.~Holzner},
\newblock \bibinfo{title}{Shear-scaling-based approach for irreversible energy loss estimation in stenotic aortic flow--an in vitro study},
\newblock \bibinfo{journal}{Journal of biomechanics} \bibinfo{volume}{56} (\bibinfo{year}{2017}) \bibinfo{pages}{89--96}.
%Type = Article
\bibitem[{Zeugin et~al.(2024)Zeugin, Coulter, G{\"u}lan, Studart, and Holzner}]{zeugin2024vitro}
\bibinfo{author}{T.~Zeugin}, \bibinfo{author}{F.~B. Coulter}, \bibinfo{author}{U.~G{\"u}lan}, \bibinfo{author}{A.~R. Studart}, \bibinfo{author}{M.~Holzner},
\newblock \bibinfo{title}{In vitro investigation of the blood flow downstream of a 3d-printed aortic valve},
\newblock \bibinfo{journal}{Scientific Reports} \bibinfo{volume}{14} (\bibinfo{year}{2024}) \bibinfo{pages}{1572}.
%Type = Misc
\bibitem[{{MicroPort}(1 26)}]{thickness_av}
\bibinfo{author}{{MicroPort}}, \bibinfo{title}{Severe aortic valve stenosis}, \bibinfo{howpublished}{\url{https://microport.com/patients/severe-aortic-valve-stenosis}}, \bibinfo{year}{2024-01-26}.
%Type = Article
\bibitem[{Stein and Sabbah(1976)}]{stein1976turbulent}
\bibinfo{author}{P.~D. Stein}, \bibinfo{author}{H.~N. Sabbah},
\newblock \bibinfo{title}{Turbulent blood flow in the ascending aorta of humans with normal and diseased aortic valves.},
\newblock \bibinfo{journal}{Circulation research} \bibinfo{volume}{39} (\bibinfo{year}{1976}) \bibinfo{pages}{58--65}.
%Type = Article
\bibitem[{Vreman(2004)}]{vreman2004eddy}
\bibinfo{author}{A.~Vreman},
\newblock \bibinfo{title}{An eddy-viscosity subgrid-scale model for turbulent shear flow: Algebraic theory and applications},
\newblock \bibinfo{journal}{Physics of fluids} \bibinfo{volume}{16} (\bibinfo{year}{2004}) \bibinfo{pages}{3670--3681}.
%Type = Article
\bibitem[{Pal et~al.(2014)Pal, Anupindi, Delorme, Ghaisas, Shetty, and Frankel}]{pal2014large}
\bibinfo{author}{A.~Pal}, \bibinfo{author}{K.~Anupindi}, \bibinfo{author}{Y.~Delorme}, \bibinfo{author}{N.~Ghaisas}, \bibinfo{author}{D.~A. Shetty}, \bibinfo{author}{S.~H. Frankel},
\newblock \bibinfo{title}{Large eddy simulation of transitional flow in an idealized stenotic blood vessel: evaluation of subgrid scale models},
\newblock \bibinfo{journal}{Journal of biomechanical engineering} \bibinfo{volume}{136} (\bibinfo{year}{2014}) \bibinfo{pages}{071009}.
%Type = Article
\bibitem[{Katz et~al.(2023)Katz, Caiazzo, Moreau, Wilbrandt, Br{\"u}ning, Goubergrits, and John}]{katz2023impact}
\bibinfo{author}{S.~Katz}, \bibinfo{author}{A.~Caiazzo}, \bibinfo{author}{B.~Moreau}, \bibinfo{author}{U.~Wilbrandt}, \bibinfo{author}{J.~Br{\"u}ning}, \bibinfo{author}{L.~Goubergrits}, \bibinfo{author}{V.~John},
\newblock \bibinfo{title}{Impact of turbulence modeling on the simulation of blood flow in aortic coarctation},
\newblock \bibinfo{journal}{International Journal for Numerical Methods in Biomedical Engineering} \bibinfo{volume}{39} (\bibinfo{year}{2023}) \bibinfo{pages}{e3695}.
%Type = Article
\bibitem[{Manchester et~al.(2021)Manchester, Pirola, Salmasi, O’Regan, Athanasiou, and Xu}]{manchester2021analysis}
\bibinfo{author}{E.~L. Manchester}, \bibinfo{author}{S.~Pirola}, \bibinfo{author}{M.~Y. Salmasi}, \bibinfo{author}{D.~P. O’Regan}, \bibinfo{author}{T.~Athanasiou}, \bibinfo{author}{X.~Y. Xu},
\newblock \bibinfo{title}{Analysis of turbulence effects in a patient-specific aorta with aortic valve stenosis},
\newblock \bibinfo{journal}{Cardiovascular engineering and technology} \bibinfo{volume}{12} (\bibinfo{year}{2021}) \bibinfo{pages}{438--453}.
%Type = Book
\bibitem[{Belytschko et~al.(2014)Belytschko, Liu, Moran, and Elkhodary}]{belytschko2014nonlinear}
\bibinfo{author}{T.~Belytschko}, \bibinfo{author}{W.~K. Liu}, \bibinfo{author}{B.~Moran}, \bibinfo{author}{K.~Elkhodary}, \bibinfo{title}{Nonlinear finite elements for continua and structures}, \bibinfo{publisher}{John wiley \& sons}, \bibinfo{year}{2014}.
%Type = Article
\bibitem[{De~Hart et~al.(2003)De~Hart, Peters, Schreurs, and Baaijens}]{de2003three}
\bibinfo{author}{J.~De~Hart}, \bibinfo{author}{G.~Peters}, \bibinfo{author}{P.~Schreurs}, \bibinfo{author}{F.~Baaijens},
\newblock \bibinfo{title}{A three-dimensional computational analysis of fluid--structure interaction in the aortic valve},
\newblock \bibinfo{journal}{Journal of biomechanics} \bibinfo{volume}{36} (\bibinfo{year}{2003}) \bibinfo{pages}{103--112}.
%Type = Article
\bibitem[{Lehmkuhl et~al.(2019)Lehmkuhl, Houzeaux, Owen, Chrysokentis, and Rodr{\'\i}guez}]{lehmkuhl2019low}
\bibinfo{author}{O.~Lehmkuhl}, \bibinfo{author}{G.~Houzeaux}, \bibinfo{author}{H.~Owen}, \bibinfo{author}{G.~Chrysokentis}, \bibinfo{author}{I.~Rodr{\'\i}guez},
\newblock \bibinfo{title}{A low-dissipation finite element scheme for scale resolving simulations of turbulent flows},
\newblock \bibinfo{journal}{Journal of Computational Physics} \bibinfo{volume}{390} (\bibinfo{year}{2019}) \bibinfo{pages}{51--65}.
%Type = Article
\bibitem[{V{\'a}zquez et~al.(2016)V{\'a}zquez, Houzeaux, Koric, Artigues, Aguado-Sierra, Ar{\'\i}s, Mira, Calmet, Cucchietti, Owen et~al.}]{vazquez2016alya}
\bibinfo{author}{M.~V{\'a}zquez}, \bibinfo{author}{G.~Houzeaux}, \bibinfo{author}{S.~Koric}, \bibinfo{author}{A.~Artigues}, \bibinfo{author}{J.~Aguado-Sierra}, \bibinfo{author}{R.~Ar{\'\i}s}, \bibinfo{author}{D.~Mira}, \bibinfo{author}{H.~Calmet}, \bibinfo{author}{F.~Cucchietti}, \bibinfo{author}{H.~Owen}, et~al.,
\newblock \bibinfo{title}{Alya: Multiphysics engineering simulation toward exascale},
\newblock \bibinfo{journal}{Journal of computational science} \bibinfo{volume}{14} (\bibinfo{year}{2016}) \bibinfo{pages}{15--27}.
%Type = Article
\bibitem[{Santiago et~al.(2018)Santiago, Aguado-Sierra, Zavala-Ak{\'e}, Doste-Beltran, G{\'o}mez, Ar{\'\i}s, Cajas, Casoni, and V{\'a}zquez}]{santiago2018fully}
\bibinfo{author}{A.~Santiago}, \bibinfo{author}{J.~Aguado-Sierra}, \bibinfo{author}{M.~Zavala-Ak{\'e}}, \bibinfo{author}{R.~Doste-Beltran}, \bibinfo{author}{S.~G{\'o}mez}, \bibinfo{author}{R.~Ar{\'\i}s}, \bibinfo{author}{J.~C. Cajas}, \bibinfo{author}{E.~Casoni}, \bibinfo{author}{M.~V{\'a}zquez},
\newblock \bibinfo{title}{Fully coupled fluid-electro-mechanical model of the human heart for supercomputers},
\newblock \bibinfo{journal}{International journal for numerical methods in biomedical engineering} \bibinfo{volume}{34} (\bibinfo{year}{2018}) \bibinfo{pages}{e3140}.
%Type = Article
\bibitem[{T{\"o}rnqvist et~al.(1985)T{\"o}rnqvist, Vartia, and Vartia}]{tornqvist1985should}
\bibinfo{author}{L.~T{\"o}rnqvist}, \bibinfo{author}{P.~Vartia}, \bibinfo{author}{Y.~O. Vartia},
\newblock \bibinfo{title}{How should relative changes be measured?},
\newblock \bibinfo{journal}{The American Statistician} \bibinfo{volume}{39} (\bibinfo{year}{1985}) \bibinfo{pages}{43--46}.
%Type = Article
\bibitem[{Zingaro et~al.(2021)Zingaro, Dede', Menghini, Quarteroni et~al.}]{zingaro2021hemodynamics}
\bibinfo{author}{A.~Zingaro}, \bibinfo{author}{L.~Dede'}, \bibinfo{author}{F.~Menghini}, \bibinfo{author}{A.~Quarteroni}, et~al.,
\newblock \bibinfo{title}{Hemodynamics of the heart’s left atrium based on a variational multiscale-les numerical method},
\newblock \bibinfo{journal}{European Journal of Mechanics-B/Fluids} \bibinfo{volume}{89} (\bibinfo{year}{2021}) \bibinfo{pages}{380--400}.
%Type = Article
\bibitem[{Bonow et~al.(1998)Bonow, Carabello, De~Leon, Edmunds~Jr, Fedderly, Freed, Gaasch, McKay, Nishimura, O'Gara et~al.}]{bonow1998acc}
\bibinfo{author}{R.~Bonow}, \bibinfo{author}{B.~Carabello}, \bibinfo{author}{A.~De~Leon}, \bibinfo{author}{L.~Edmunds~Jr}, \bibinfo{author}{B.~Fedderly}, \bibinfo{author}{M.~Freed}, \bibinfo{author}{W.~Gaasch}, \bibinfo{author}{C.~McKay}, \bibinfo{author}{R.~Nishimura}, \bibinfo{author}{P.~O'Gara}, et~al.,
\newblock \bibinfo{title}{Acc/aha guidelines for the management of patients with valvular heart disease. executive summary. a report of the american college of cardiology/american heart association task force on practice guidelines (committee on management of patients with valvular heart disease)},
\newblock \bibinfo{journal}{The Journal of heart valve disease} \bibinfo{volume}{7} (\bibinfo{year}{1998}) \bibinfo{pages}{672--707}.
%Type = Article
\bibitem[{Zhang et~al.(2020)Zhang, Chen, Griffith, and Wu}]{zhang2020computational}
\bibinfo{author}{J.~Zhang}, \bibinfo{author}{Z.~Chen}, \bibinfo{author}{B.~P. Griffith}, \bibinfo{author}{Z.~J. Wu},
\newblock \bibinfo{title}{Computational characterization of flow and blood damage potential of the new maglev ch-vad pump versus the hvad and heartmate ii pumps},
\newblock \bibinfo{journal}{The International journal of artificial organs} \bibinfo{volume}{43} (\bibinfo{year}{2020}) \bibinfo{pages}{653--662}.
%Type = Article
\bibitem[{Keele(1952)}]{keele1952leonardo}
\bibinfo{author}{K.~D. Keele},
\newblock \bibinfo{title}{Leonardo da vinci as physiologist},
\newblock \bibinfo{journal}{Postgraduate medical journal} \bibinfo{volume}{28} (\bibinfo{year}{1952}) \bibinfo{pages}{521}.
%Type = Article
\bibitem[{Robicsek(1991)}]{robicsek1991leonardo}
\bibinfo{author}{F.~Robicsek},
\newblock \bibinfo{title}{Leonardo da vinci and the sinuses of valsalva},
\newblock \bibinfo{journal}{The Annals of thoracic surgery} \bibinfo{volume}{52} (\bibinfo{year}{1991}) \bibinfo{pages}{328--335}.
%Type = Article
\bibitem[{Chnafa et~al.(2014)Chnafa, Mendez, and Nicoud}]{chnafa2014image}
\bibinfo{author}{C.~Chnafa}, \bibinfo{author}{S.~Mendez}, \bibinfo{author}{F.~Nicoud},
\newblock \bibinfo{title}{Image-based large-eddy simulation in a realistic left heart},
\newblock \bibinfo{journal}{Computers \& Fluids} \bibinfo{volume}{94} (\bibinfo{year}{2014}) \bibinfo{pages}{173--187}.
%Type = Article
\bibitem[{Otto et~al.(2021)Otto, Nishimura, Bonow, Carabello, Erwin, Gentile, Jneid, Krieger, Mack, McLeod et~al.}]{otto2021acc}
\bibinfo{author}{C.~Otto}, \bibinfo{author}{R.~Nishimura}, \bibinfo{author}{R.~Bonow}, \bibinfo{author}{B.~Carabello}, \bibinfo{author}{J.~Erwin}, \bibinfo{author}{F.~Gentile}, \bibinfo{author}{H.~Jneid}, \bibinfo{author}{E.~Krieger}, \bibinfo{author}{M.~Mack}, \bibinfo{author}{C.~McLeod}, et~al.,
\newblock \bibinfo{title}{Acc/aha guideline for the management of patients with valvular heart disease: Executive summary: A report of the american college of cardiology/american heart association joint committee on clinical practice guidelines},
\newblock \bibinfo{journal}{Circulation} \bibinfo{volume}{143} (\bibinfo{year}{2021}) \bibinfo{pages}{e35--e71}.
%Type = Article
\bibitem[{Baumgartner et~al.(2017)Baumgartner, Falk, Bax, De~Bonis, Hamm, Holm, Iung, Lancellotti, Lansac, Rodriguez et~al.}]{baumgartner20172017}
\bibinfo{author}{H.~Baumgartner}, \bibinfo{author}{V.~Falk}, \bibinfo{author}{J.~J. Bax}, \bibinfo{author}{M.~De~Bonis}, \bibinfo{author}{C.~Hamm}, \bibinfo{author}{P.~J. Holm}, \bibinfo{author}{B.~Iung}, \bibinfo{author}{P.~Lancellotti}, \bibinfo{author}{E.~Lansac}, \bibinfo{author}{D.~Rodriguez}, et~al.,
\newblock \bibinfo{title}{2017 {ESC/EACTS} guidelines for the management of valvular heart disease},
\newblock \bibinfo{journal}{European heart journal} \bibinfo{volume}{38} (\bibinfo{year}{2017}) \bibinfo{pages}{2739--2786}.
%Type = Article
\bibitem[{Rossini et~al.(2016)Rossini, Martinez-Legazpi, Vu, Fernandez-Friera, Del~Villar, Rodriguez-Lopez, Benito, Borja, Pastor-Escuredo, Yotti et~al.}]{rossini2016clinical}
\bibinfo{author}{L.~Rossini}, \bibinfo{author}{P.~Martinez-Legazpi}, \bibinfo{author}{V.~Vu}, \bibinfo{author}{L.~Fernandez-Friera}, \bibinfo{author}{C.~P. Del~Villar}, \bibinfo{author}{S.~Rodriguez-Lopez}, \bibinfo{author}{Y.~Benito}, \bibinfo{author}{M.-G. Borja}, \bibinfo{author}{D.~Pastor-Escuredo}, \bibinfo{author}{R.~Yotti}, et~al.,
\newblock \bibinfo{title}{A clinical method for mapping and quantifying blood stasis in the left ventricle},
\newblock \bibinfo{journal}{Journal of biomechanics} \bibinfo{volume}{49} (\bibinfo{year}{2016}) \bibinfo{pages}{2152--2161}.
%Type = Article
\bibitem[{Hatoum et~al.(2019)Hatoum, Dollery, Lilly, Crestanello, and Dasi}]{hatoum2019impact}
\bibinfo{author}{H.~Hatoum}, \bibinfo{author}{J.~Dollery}, \bibinfo{author}{S.~M. Lilly}, \bibinfo{author}{J.~Crestanello}, \bibinfo{author}{L.~P. Dasi},
\newblock \bibinfo{title}{Impact of patient-specific morphologies on sinus flow stasis in transcatheter aortic valve replacement: an in vitro study},
\newblock \bibinfo{journal}{The Journal of thoracic and cardiovascular surgery} \bibinfo{volume}{157} (\bibinfo{year}{2019}) \bibinfo{pages}{540--549}.
%Type = Article
\bibitem[{Huang et~al.(2019)Huang, Wang, Lin, Li, Shan, and Zheng}]{huang2019comparison}
\bibinfo{author}{J.~Huang}, \bibinfo{author}{Y.~Wang}, \bibinfo{author}{L.~Lin}, \bibinfo{author}{Z.~Li}, \bibinfo{author}{Z.~Shan}, \bibinfo{author}{S.~Zheng},
\newblock \bibinfo{title}{Comparison of dynamic changes in aortic diameter during the cardiac cycle measured by computed tomography angiography and transthoracic echocardiography},
\newblock \bibinfo{journal}{Journal of Vascular Surgery} \bibinfo{volume}{69} (\bibinfo{year}{2019}) \bibinfo{pages}{1538--1544}.
%Type = Article
\bibitem[{Binter et~al.(2016)Binter, G{\"u}lan, Holzner, and Kozerke}]{binter2016accuracy}
\bibinfo{author}{C.~Binter}, \bibinfo{author}{U.~G{\"u}lan}, \bibinfo{author}{M.~Holzner}, \bibinfo{author}{S.~Kozerke},
\newblock \bibinfo{title}{On the accuracy of viscous and turbulent loss quantification in stenotic aortic flow using phase-contrast mri},
\newblock \bibinfo{journal}{Magnetic resonance in medicine} \bibinfo{volume}{76} (\bibinfo{year}{2016}) \bibinfo{pages}{191--196}.
%Type = Article
\bibitem[{Knobloch et~al.(2014)Knobloch, Binter, G{\"u}lan, Sigfridsson, Holzner, L{\"u}thi, and Kozerke}]{knobloch2014mapping}
\bibinfo{author}{V.~Knobloch}, \bibinfo{author}{C.~Binter}, \bibinfo{author}{U.~G{\"u}lan}, \bibinfo{author}{A.~Sigfridsson}, \bibinfo{author}{M.~Holzner}, \bibinfo{author}{B.~L{\"u}thi}, \bibinfo{author}{S.~Kozerke},
\newblock \bibinfo{title}{Mapping mean and fluctuating velocities by bayesian multipoint mr velocity encoding-validation against 3d particle tracking velocimetry},
\newblock \bibinfo{journal}{Magnetic resonance in medicine} \bibinfo{volume}{71} (\bibinfo{year}{2014}) \bibinfo{pages}{1405--1415}.
%Type = Article
\bibitem[{Petersson et~al.(2012)Petersson, Dyverfeldt, and Ebbers}]{petersson2012assessment}
\bibinfo{author}{S.~Petersson}, \bibinfo{author}{P.~Dyverfeldt}, \bibinfo{author}{T.~Ebbers},
\newblock \bibinfo{title}{Assessment of the accuracy of mri wall shear stress estimation using numerical simulations},
\newblock \bibinfo{journal}{Journal of Magnetic Resonance Imaging} \bibinfo{volume}{36} (\bibinfo{year}{2012}) \bibinfo{pages}{128--138}.
%Type = Article
\bibitem[{Liu et~al.(2020)Liu, Yang, Lan, and Marsden}]{liu2020fluid}
\bibinfo{author}{J.~Liu}, \bibinfo{author}{W.~Yang}, \bibinfo{author}{I.~S. Lan}, \bibinfo{author}{A.~L. Marsden},
\newblock \bibinfo{title}{Fluid-structure interaction modeling of blood flow in the pulmonary arteries using the unified continuum and variational multiscale formulation},
\newblock \bibinfo{journal}{Mechanics research communications} \bibinfo{volume}{107} (\bibinfo{year}{2020}) \bibinfo{pages}{103556}.
%Type = Article
\bibitem[{Cai et~al.(2021)Cai, Zhang, Li, Zhu, Luo, and Gao}]{cai2021comparison}
\bibinfo{author}{L.~Cai}, \bibinfo{author}{R.~Zhang}, \bibinfo{author}{Y.~Li}, \bibinfo{author}{G.~Zhu}, \bibinfo{author}{X.~Luo}, \bibinfo{author}{H.~Gao},
\newblock \bibinfo{title}{The comparison of different constitutive laws and fiber architectures for the aortic valve on fluid--structure interaction simulation},
\newblock \bibinfo{journal}{Frontiers in Physiology} \bibinfo{volume}{12} (\bibinfo{year}{2021}) \bibinfo{pages}{682893}.
%Type = Article
\bibitem[{Criseo et~al.(2024)Criseo, Fumagalli, Quarteroni, Marianeschi, and Vergara}]{criseo2024computational}
\bibinfo{author}{M.~Criseo}, \bibinfo{author}{I.~Fumagalli}, \bibinfo{author}{A.~Quarteroni}, \bibinfo{author}{S.~M. Marianeschi}, \bibinfo{author}{C.~Vergara},
\newblock \bibinfo{title}{Computational haemodynamics for pulmonary valve replacement by means of a reduced fluid-structure interaction model},
\newblock \bibinfo{journal}{MOX Report (preprint)} \bibinfo{volume}{01} (\bibinfo{year}{2024}).

\end{thebibliography}
